\documentclass[fleqn,usenatbib]{mnras}
\usepackage{newtxtext,newtxmath}

\usepackage[T1]{fontenc}
\usepackage{ae,aecompl}

\usepackage{graphicx}	
\usepackage{amsmath}	
\usepackage{amssymb}	
\usepackage{siunitx}
\usepackage{xspace}
\usepackage{esvect}
\usepackage{subfigure}
\usepackage{etoolbox}
\makeatletter
\patchcmd\@combinedblfloats{\box\@outputbox}{\unvbox\@outputbox}{}{%
   \errmessage{\noexpand\@combinedblfloats could not be patched}%
}%
 \makeatother
 
\newcommand{\curL}{\mathcal{L}}
\newcommand{\curH}{\mathcal{H}}
\newcommand{\curM}{\mathcal{M}}

\newcommand{\Rstar}{\ensuremath{R_{\star}}\xspace}
\newcommand{\msini}{\ensuremath{m \sin I}\xspace} 
\newcommand{\ms}{\ensuremath{\mathrm{m}\,\mathrm{s}^{-1}}\xspace}
\newcommand{\masyr}{\ensuremath{\mathrm{mas}\,\mathrm{yr}^{-1}}\xspace}
\newcommand{\Me}{\ensuremath{M_{\oplus}}\xspace} 
\renewcommand{\Re}{\ensuremath{R_{\oplus}}\xspace}

\newcommand{\Mj}{\ensuremath{M_{\rm{Jup}}}\xspace} 

\newcommand{\Rsun}{\ensuremath{R_{\odot}}\xspace}
\newcommand{\Msun}{\ensuremath{M_{\odot}}\xspace}
\newcommand{\Lone}{\ensuremath{\boldsymbol{\hat{L}}_1}}
\newcommand{\Ltwo}{\ensuremath{\boldsymbol{\hat{L}}_2}}
\newcommand{\Lstar}{\ensuremath{\boldsymbol{\hat{L}}_\star}}
\newcommand{\Lppd}{\ensuremath{\boldsymbol{\hat{L}}_{\rm{ppd}}}}
\newcommand{\Ldd}{\ensuremath{\boldsymbol{\hat{L}}_{\rm{dd}}}}

\newcommand{\Lonei}{\ensuremath{\boldsymbol{\hat{L}}_{\rm{1,i}}}}
\newcommand{\Ltwoi}{\ensuremath{\boldsymbol{\hat{L}}_{\rm{2,i}}}}
\newcommand{\Lstari}{\ensuremath{\boldsymbol{\hat{L}}_{\rm{\star,i}}}}
\newcommand{\Lppdi}{\ensuremath{\boldsymbol{\hat{L}}_{\rm{ppd,i}}}}
\newcommand{\Lddi}{\ensuremath{\boldsymbol{\hat{L}}_{\rm{dd,i}}}}

\newcommand{\Lonelt}{\ensuremath{\boldsymbol{\hat{L}}_{\rm{1,lt}}}}
\newcommand{\Ltwolt}{\ensuremath{\boldsymbol{\hat{L}}_{\rm{2,lt}}}}
\newcommand{\Lstarlt}{\ensuremath{\boldsymbol{\hat{L}}_{\rm{\star,lt}}}}

\newcommand{\Lddlt}{\ensuremath{\boldsymbol{\hat{L}}_{\rm{dd,lt}}}}

\title[{\bf Mutual inclination between $\pi$ Men b and c}]{Evidence for a high mutual inclination between the cold Jupiter and transiting super Earth orbiting $\pi$ Men}

\author[Xuan \& Wyatt]{
Jerry W. Xuan,\thanks{E-mail: wx239@cam.ac.uk} and
Mark C. Wyatt
\\
Institute of Astronomy, University of Cambridge, Madingley Road, Cambridge CB3 0HA, UK
}

\date{Accepted 2020 July 2. Received 2020 June 4; in original form 2020 April 1}

\pubyear{2020}

\begin{document}
\label{firstpage}
\pagerange{\pageref{firstpage}--\pageref{lastpage}}
\maketitle

\begin{abstract}

$\pi$ Men hosts a transiting super Earth ($P\approx6.27$~d, $m\approx4.82$~\Me, $R\approx2.04~\Re$) discovered by TESS and a cold Jupiter ($P\approx2093$~d, $\msini\approx10.02$~\Mj, $e\approx0.64$) discovered from radial velocity. We use {\it Gaia} DR2 and {\it Hipparcos} astrometry to derive the star's velocity caused by the orbiting planets and constrain the cold Jupiter's sky-projected inclination ($I_b=41-65\degr$). From this we derive the mutual inclination ($\Delta I$) between the two planets, and find that $49\degr < \Delta I < 131\degr$ (1$\sigma$), and $28\degr < \Delta I < 152\degr$ (2$\sigma$). We examine the dynamics of the system using $N$-body simulations, and find that potentially large oscillations in the super Earth's eccentricity and inclination are suppressed by general relativistic precession. However, nodal precession of the inner orbit around the invariable plane causes the super Earth to only transit between 7-22 per cent of the time, and to usually be observed as misaligned with the stellar spin axis. We repeat our analysis for HAT-P-11, finding a large $\Delta I$ between its close-in Neptune and cold Jupiter and similar dynamics. $\pi$ Men and HAT-P-11 are prime examples of systems where dynamically hot outer planets excite their inner planets, with the effects of increasing planet eccentricities, planet-star misalignments, and potentially reducing the transit multiplicity. Formation of such systems likely involves scattering between multiple giant planets or misaligned protoplanetary discs. Future imaging of the faint debris disc in $\pi$ Men and precise constraints on its stellar spin orientation would provide strong tests for these formation scenarios.

\end{abstract}


\section{Introduction}

Although important for understanding planetary dynamics and evolution, the mutual inclination ($\Delta I$) between neighbouring exoplanets has been one of the more difficult orbital parameters to measure. The transit and radial velocity techniques do not naturally yield all the information necessary for measuring $\Delta I$, namely the sky-projected inclinations and positions of ascending nodes of the orbits. Therefore, most measurements of $\Delta I$ have relied on either statistical arguments, or exceptional cases.

On the statistical side, constraints on $\Delta I$ have been made by analysing the transit duration ratios of adjacent planets in {\it Kepler} systems \citep{fabrycky_architecture_2014}, in which case $\Delta I$ is found to be in the range of $1-2\degr$, except in systems with ultra-short period planets, in which case $\Delta I \gtrsim7\degr$ \citep{dai_larger_2018-1}. On the other hand, there is indirect evidence of a preference for $\Delta I \approx40\degr$ in systems with one warm Jupiter ($0.1 < a < 1$ au) and one cold Jupiter ($a >1$ au), based on the tendency for their periastra to be misaligned by $\sim 90\degr$ \citep{dawson_class_2014}.

In systems with strongly interacting planets, direct constraints on $\Delta I$ have been obtained from periodic variations in the transit times or transit durations of planets \citep[e.g.][]{nesvorny_detection_2012, sanchis-ojeda_alignment_2012, mills_kepler-108_2017}, where $\Delta I$ is usually found to be $\lesssim 10\degr$, with the exception of {\it Kepler}-108, which has two Saturn-mass planets inclined by $16-35\degr$. A rare case of a planet-planet transit in {\it Kepler}-89 has also been used to directly measure $\Delta I$, which is found to be $\sim 1\degr$ between the two eclipsing planets \citep{hirano_planet-planet_2012}. More directly, \citet{mcarthur_new_2010} combined RV with astrometric monitoring of the host star to find $\Delta I\approx30\degr$ between two giant planets in the ups And system.

Almost all previous measurements on $\Delta I$ were made between planets of similar masses. It is also important to determine the mutual inclinations between planets with different masses and even different classes, the two most abundant of which are super Earths (SEs) and cold Jupiters (CJs). Broadly speaking, SEs can be defined as planets with $1-4~\Re$ or $1-10~\Me$ \citep{bryan_excess_2019}, orbiting with periods less than 1 yr, but typically on the order of tens of days -- the classic {\it Kepler} planet. CJs are gas giant planets with periods longer than 1 yr, typically discovered with RV. The typical mass ratio between SEs and CJs is $\lesssim 1$ per cent, and these systems are hierarchical (i.e. $a_{\rm{out}} \gg a_{\rm{in}}$).

Interestingly, both \citet{zhu_super_2018} and \citet{bryan_excess_2019} found that SEs and CJs are statistically common around each other, with SEs accompanied by CJs around $30-40$ per cent of the time, and CJs accompanied by SEs almost all the time. It is unclear why such a correlation exists. Indeed, the strong correlation directly contradicts certain formation theories which postulate that outer CJs act as barriers to forming SEs by reducing pebble flux \citep{ormel_formation_2017} or hindering inward migration of SEs that originally formed farther out \citep{izidoro_gas_2015}. In order to study the dynamics in SE+CJ systems and shed light on their formation processes, it is necessary to have a complete picture of their 3D orbital architectures. In particular, knowledge of $\Delta I$ between the SE and CJ orbits would provide crucial constraints on the dynamics.

Toward this effort, \citet{masuda_mutual_2020} used a sample of three {\it Kepler} systems with transiting SEs and transiting CJs to statistically derive a $1\sigma$ interval of $11.8\degr^{+12.7\degr}_{-5.5\degr}$ for the mutual inclination distribution between inner SEs and outer CJs in the {\it Kepler} sample. Rather than direct measurements, their analysis was based on geometric arguments about the average transit probability of CJs in the entire sample of {\it Kepler} systems that host SEs. As such, no direct measurements exist for individual SE+CJ systems, precluding a more detailed examination of the dynamics.

In this paper, we perform a direct measurement of mutual inclination for the SE+CJ system $\pi$ Men, which hosts a transiting SE ($P\approx6.27$ d, $m\approx4.82$~\Me, $R\approx2.04~\Re$) discovered by TESS \citep{huang_tess_2018}, and an eccentric outer CJ ($P\approx2093$ d, $\msini\approx10.02$~\Mj. $e\approx0.64$) discovered with RV \citep{jones_probable_2002-1}. In $\pi$ Men, the inclination is measured for the transiting SE \citep{huang_tess_2018}. To constrain the inclination and ascending node of the outer CJ, we analyse proper motion data from a combination of {\it Gaia} DR2 \citep{gaia_collaboration_gaia_2016, gaia_collaboration_2018} and {\it Hipparcos} astrometry \citep{perryman_hipparcos_1997, van_leeuwen_validation_2007} and perform a joint fit of the astrometric and RV data. Despite the unknown ascending node of the SE, we are able to derive a relatively strong constraint on the mutual inclination. In the end, we find that $\approx85$ per cent of the possible $\Delta I$ are between $39.2-140.8\degr$, which is the regime where large amplitude oscillations in eccentricity and inclination can take place \citep{kozai_secular_1962, lidov_evolution_1962}. This motivates us to carefully examine the dynamics at play in $\pi$ Men using both analytic considerations and $N$-body simulations.

The technique of combining {\it Gaia} DR2 and {\it Hipparcos} was described in \citet{brandt_hipparcos-gaia_2018} and \citet{kervella_stellar_2019}, and has been successfully applied to brown dwarf companions \citep[e.g.][]{brandt_precise_2019, brandt_dynamical_2019, dupuy_model_2019}, a nearby CJ orbting $\epsilon$ Indi A \citep{feng_detection_2019}, and the planet candidate Proxima Centauri c \citep{kervella_orbital_2020}. Using the example of $\pi$ Men, we show that this technique can be applied to similar planetary systems with a large astrometric signal in order to unveil the dynamics at play. Indeed, we also apply our method to HAT-P-11, and confirm the postulation in \citet{yee_hat-p-11_2018} that the two planets in HAT-P-11 might have $\Delta I \gtrsim50\degr$, although the astrometric data for HAT-P-11 is less significant than that for $\pi$ Men. HAT-P-11 is a hierarchical system with an inner Neptune and an outer CJ. Although it does not perfectly fit the SE+CJ framework, it is a similar dynamical system to $\pi$ Men, allowing us to draw comparisons between the two systems. As new TESS planets are discovered, and RV follow-ups of {\it Kepler} and TESS planets continue, more systems like $\pi$ Men can be studied in such a way.

This paper is structured as follows. In \S\ref{sec:basic_method}, we explain how astrometric differences between {\it Gaia} DR2 and {\it Hipparcos} can be used as tangential velocities and combined with radial velocity data to constrain orbits. We then present simple and detailed orbit fits for a brown dwarf validation target, $\pi$ Men, and HAT-P-11 in \S\ref{sec:orbit_fits}, and derive a distribution of the mutual inclination between $\pi$ Men b and c, which we find to be large. A similarly large mutual inclination is also found between HAT-P-11 b and c. In \S\ref{sec:analytic_dynamics}, we summarize the secular dynamics relevant in $\pi$ Men and HAT-P-11. Numerical simulations are used \S\ref{sec:simulations} to thoroughly examine the dynamics in $\pi$ Men. Based on the simulations, we demonstrate in \S\ref{sec:nod_prec_effects} that nodal precession of the inner orbit has interesting effects on the long-term transit probability of $\pi$ Men c and its alignment relative to the stellar spin axis. We generalize our findings with previous results for HAT-P-11, which exhibits similar dynamical effects as $\pi$ Men. In \S\ref{sec:discuss}, we discuss the implications of our findings for the formation and evolution of {these} systems. Finally, we conclude in \S\ref{sec:conclusion}.

\section{The three-component stellar velocity} \label{sec:basic_method}

\subsection{Equations of motion}\label{sec:eq_motion}

Consider a two-body system composed of a star of mass $M_{\star}$ and a planet (or companion) of mass $m$. In the orbital plane, the relative position between the planet and the star is defined by the orbital elements $a$, $e$, and $f$, where $a$ is the semi-major axis, $e$ is the eccentricity, and $f$ is the true anomaly. 

In our astrometric and RV measurements, we observe the motion of the star around the barycentre. The orbital distance between the star and the barycentre is given by
\begin{equation}
    r_\star = \left(\frac{m}{M_\star+m}\right)\frac{a(1-e^2)}{1+e\cos{f}}.
    \label{eq:r_star}
\end{equation}
We define the stellar orbit in Cartesian coordinates ($x, y, z$) centred at the barycentre as follows: $x$ points toward the pericentre of the orbit, $y$ lies in the orbital plane perpendicular to $x$, and $z$ points along the direction of angular momentum to complete the right-handed triad. The position vector for the star in the orbital plane is then given by
\begin{equation}\label{eq:xyz}
\begin{aligned}
    x_\star &= r_\star\cos{f}, \\
    y_\star &= r_\star\sin{f}, \\
    z_\star &= 0.
\end{aligned}
\end{equation}
For the observer, we define a sky-plane coordinate system ($X, Y, Z$) also centred at the barycentre, with $X$ pointing in the direction of declination ($\delta$), $Y$ pointing in the direction of right-ascension ($\alpha$), and $Z$ pointing {\it toward} the observer to complete the right-handed triad. We stress that this is different from the radial velocity convention, where $+Z$ points away from the observer \citep[e.g. in][]{fulton_radvel_2018}. We adopt our coordinate system in order to be consistent with the astrometry convention that $+X = \delta$ and $+Y = \alpha$ \citep[e.g. in][]{feng_detection_2019,pearce_orbital_2020}. Transforming $x_\star$, $y_\star$, and $z_\star$ into the sky plane involves three rotations about the orbital plane through the angles $I$, $\omega_\star$, and $\Omega$. Here, $I$ is the inclination of the orbit with respect to the reference frame, $\Omega$ is the longitude of ascending node, and $\omega_\star$ is the argument of periastron of the star.

Due to the differing notations in the literature, we wish to specify our definitions for the angles and directions. First, we refer to the stellar orbit, and use $\omega_\star$ instead of $\omega_p$; the two angles are out of phase by $\pi$, and it is not always clear which is being quoted in the literature. The remaining angles $I$, $\Omega$, and $f$ are identical for both star and companion. The inclination ($I$) is defined as the angle between the angular momentum vector of the orbit and the $+Z$ axis. The ascending node refers to the node where the object moves in the $+Z$ direction, and the longitude of the ascending node ($\Omega$) is measured eastward from the reference direction ($+X$ or $\delta$) to the ascending node. Finally, $\omega_\star$ is measured from the ascending node to the stellar periastron, in the direction of the orbit.

With these definitions in mind, we apply the coordinate rotations to Eq.~\ref{eq:xyz}, and get the stellar position vector in the sky plane \citep[see e.g.][]{murray_solar_2000}
\begin{equation}\label{eq:positions}
\begin{aligned}
    X_\star &= r_\star[\cos{\Omega}\cos(\omega_\star+f) - \sin{\Omega}\sin(\omega_\star+f)\cos{I}], \\
    Y_\star &= r_\star[\sin{\Omega}\cos(\omega_\star+f) + \cos{\Omega}\sin(\omega_\star+f)\cos{I}], \\
    Z_\star &= r_\star[\sin(\omega_\star+f)\sin{I}].
\end{aligned}
\end{equation}
We can differentiate these positions with respect to the true anomaly $f$ to obtain the orbital velocities of the star in the sky plane. We denote the stellar orbital velocity $v$. Moving forward, we omit the subscript $\star$ on the velocities as we only consider the stellar orbit. After differentiating, we get
\begin{equation}
\begin{aligned}
    v_{X} &= A\,[-e\cos{\Omega}\sin{\omega_\star}-e\sin{\Omega}\cos{I}\cos{\omega_\star} \\ & -\cos{\Omega}\sin(\omega_\star+f)-\sin{\Omega}\cos{I}\cos(\omega_\star+f)],  \\
    v_{Y} &= A\,[-e\sin{\Omega}\sin{\omega_\star}+e\cos{\Omega}\cos{I}\cos{\omega_\star} \\ & -\sin{\Omega}\sin(\omega_\star+f)+\cos{\Omega}\cos{I}\cos(\omega_\star+f)], \\
    v_{Z} &= A\sin{I}\,[\cos(\omega_\star+f)+e\cos{\omega_\star}], \\
    A &= \left(\frac{m}{M_\star+m}\right)\frac{n a}{\sqrt{1-e^2}},
    \label{eq:orb_v}
\end{aligned}
\end{equation}
where we define a tangential velocity amplitude $A$ for clarity, and $n$ is the mean motion ($n=2\pi/P$), $P$ being the orbital period. Note that $A \sin{I}$ is equal to the more familiar radial velocity amplitude, $K$, which is the leading constant of $v_Z$.

In addition to the orbital motion, the star has an additional velocity due to motion of the barycentre in the galaxy. We term this velocity the systemic velocity $u$, which has three components. We denote the two tangential components as $u_{\delta}$ and $u_{\alpha}$, and the radial component as $u_{RV}$. We assume $u$ to be linear over the short time scales of observations. At any given time, the systemic velocity $u$ is added to the orbital velocity $v$. Therefore, the full stellar velocity, denoted $\mu$, is given by
\begin{equation}
\begin{aligned}
    \mu_{X} &= u_{\delta} +  v_{X}, \\
    \mu_{Y} &= u_{\alpha} + v_{Y}, \\
    \mu_{Z} &= u_{RV} + v_{Z}. \\
    \label{eq:full_velocities}
\end{aligned}
\end{equation}
With an observation of periodic trends in the radial velocity, the third component of Eq.~\ref{eq:full_velocities} can be used to constrain the orbital parameters $m \sin{I}$, $a$, $e$, $P$, $T_p$, and $\omega_\star$, where $m \sin{I}$ is the minimum companion mass and $T_p$ is the time of periastron passage. In addition, $u_{RV}$ can be measured. Therefore, if RV data are present, there are four unknowns left in Eq.~\ref{eq:orb_v}, namely $u_\delta$, $u_\alpha$, $\Omega$, and $I$ (given which $m$ can be determined). Given data for the tangential orbital velocities $v_{X}$ and $v_{Y}$, and knowledge of $u_\delta$ and $u_\alpha$, we can solve Eq.~\ref{eq:full_velocities} at a given $f$ (i.e. a given time). The true anomaly, $f$, can be calculated at the time of interest ($t$) with the orbital elements $e$, $T_p$, and $P$.

Throughout this paper, we assume a two-body system for our equations of motion. $\pi$ Men is a three-body system, but the stellar orbital velocity (Eq.~\ref{eq:orb_v}) is predominantly induced by the outer giant planet, since the velocity amplitude $A$ scales as $m a^{-1/2}$ (given $m \ll M_\star$). The ratio of velocity amplitudes between the inner and outer planets in $\pi$ Men is $\approx0.01$. For HAT-P-11, the ratio of amplitudes is $\approx0.4$ between the inner Neptune and outer giant. However, due to the short orbital period of the inner planet, we can assume that its contribution to the tangential signal is averaged to zero, as explained in \S\ref{sec:smearing} (the same also applies to $\pi$ Men).

Using equations in this section, we can use observational data for tangential and radial velocities of the star to derive the full set of orbital elements, and thereby constrain the 3D orbital architecture of planetary systems.

\subsection{Tangential velocity data:  {\it Gaia}-{\it Hipparcos} proper motion anomalies}
Our data for the tangential velocities come from the cross-calibration and comparison of the {\it Gaia} DR2 and {\it Hipparcos} catalogues, taking advantage of the long baseline between the two missions ($\approx 24.25$ yr). Specifically, \citet{brandt_hipparcos-gaia_2018} and \citet{kervella_stellar_2019} (hereafter B18 and K19) have created astrometric catalogues where proper motions and astrometric positions can be directly compared between the missions. In essence, by combining {\it Gaia} DR2 and {\it Hipparcos} astrometry, one can get three velocity vectors

\begin{itemize}
    \item {\it Hipparcos} proper motions near the epoch 1991.25, $\mu_H$
    \item {\it Gaia} DR2 proper motions near the epoch 2015.5, $\mu_G$
    \item The {\it Hipparcos}-{\it Gaia} `mean motion vector' computed from the difference between astrometric positions at the two epochs divided by the time span of 24.25 years, $\mu_{HG}$
\end{itemize}

These vectors are in units of \masyr. We follow the nomenclature of K19 in defining the `proper motion anomaly' (abbreviated PMa, symbolized $\Delta\mu$) as the difference between the proper motions and the mean motion vector. See fig. 1 in K19 for a schematic illustration of the method. From the two epochs $G$ ({\it Gaia} DR2) and $H$ ({\it Hipparcos}), we get two measurements of the PMa
\begin{equation}
\begin{aligned}
    \Delta\mu_{G} &= \mu_{G} - \mu_{HG}, \\
    \Delta\mu_{H} &= \mu_{H} - \mu_{HG}.
    \label{eq:pma_def}
\end{aligned}
\end{equation}
In reality, {\it Gaia} and {\it Hipparcos} measure positions of stars over a number of epochs, which are known as transit times. From the positions as functions of time, the astrometric measurements are extracted. Due to the coupling of the parallax motion, the intrinsic stellar motion, the reflex motion from a companion, and other systematic effects in {\it Gaia} and {\it Hipparcos}, the identification of the PMa with orbital motion is a complex task in the position space. Therefore, we chose to tackle the problem in a forward-modelling manner using velocities.

To use PMa data as tangential velocities, however, we must take into account two important systematic corrections: the proper motions $\mu_G$ and $\mu_H$ are not instantaneous velocities, but are rather averaged over the orbital arcs during the {\it Gaia} DR2 and {\it Hipparcos} observation windows; the vector $\mu_{HG}$ will contain contributions from orbital motion as the star is generally at different orbital phases at the two epochs. We explain these corrections in the following two subsections. 

\subsubsection{Correction I: Observing window averaging}\label{sec:smearing}
To understand the observing window averaging effect, we must rethink what is meant by the {\it Gaia} DR2 `epoch' or {\it Hipparcos} `epoch'. In B18, a single `characteristic' epoch is used for both {\it Gaia} DR2 and {\it Hipparcos} proper motions, and is chosen as the epoch with minimum positional uncertainty. This is more accurate than simply using 2015.5 for {\it Gaia} DR2, and 1991.25 for {\it Hipparcos}, but it is only a good approximation for objects with orbital periods significantly longer than the {\it Gaia} DR2 or {\it Hipparcos} observing periods, which is about 1227 d for {\it Hipparcos} \citep{perryman_hipparcos_1997} and 668 d for {\it Gaia} DR2 \citep{gaia_collaboration_2018}. Because the proper motions are in effect averaged over the data collection period, objects with orbital periods comparable to or shorter than the 668 d or 1227 d time scales will be subject to a smearing effect over the observing window. In an extreme case, if the orbital period is exactly 668 d, the velocities from {\it Gaia} would cancel out completely, and the PMa signal would be zero. However, even targets with longer periods will suffer some amount of smearing, and this is quantified in a statistical sense in fig 2. of K19.

For the inner planets in $\pi$ Men and HAT-P-11, $P$ is $\approx 5-6$ days, so their contribution to the measured PMa is negligible, justifying our assumption of two-body systems. On the other hand, $P \approx 6$ yr for $\pi$ Men b, and $P \approx 9$ yr for HAT-P-11 c, so orbital smearing is non-negligible for the outer planets whose orbits we wish to fit. Therefore, instead of using a single epoch, we compute the tangential velocities at every individual observation time within the {\it Gaia} DR2 and {\it Hipparcos} observing periods, and then average over them to mimic the smearing effect. Therefore, in two components, the proper motion vector at the {\it Gaia} DR2 epoch is
\begin{equation}
\begin{aligned}
    \mu_{G,\delta} &= u_{\delta} + \overline{v_{G,X}}, \\
    \mu_{G,\alpha} &= u_{\alpha} + \overline{v_{G,Y}},
    \label{eq:smear_corr}
\end{aligned}
\end{equation}
where $u$ is the systemic velocity, and $\overline{v_{G}}$ is the averaged orbital velocity over the {\it Gaia} DR2 observing window for the target. Similar equations are true for the {\it Hipparcos} epoch. 

To compute $\overline{v_G}$ and $\overline{v_H}$, we extract individual {\it Gaia} DR2 and {\it Hipparcos} transit times for our targets. The {\it Gaia} transit times can be found from the {\it Gaia} Observation Forecast Tool,\footnote{\href{https://gaia.esac.esa.int/gost/}{https://gaia.esac.esa.int/gost/}} and the {\it Hipparcos} times can be found from the Hipparcos Epoch Photometry Annex \citep{van_leeuwen_hipparcos_1997}. Assuming these transit times are all science observations, we compute expected orbital velocities for a given assumed orbit ($v_{X}$, $v_{Y}$) at each {\it Gaia} or {\it Hipparcos} transit time using Eq.~\ref{eq:orb_v}. Because {\it Gaia} DR2 and {\it Hipparcos} do not observe targets over uniform intervals, we remove observations taken within 1 day of each other to avoid giving certain epochs uneven weights. Then, we average the remaining velocities to get $\overline{v_G}$ and $\overline{v_H}$. 

\subsubsection{Correction II: Orbital phase difference}\label{sec:phase_diff}
The star is generally at different orbital positions during the {\it Gaia} and {\it Hipparcos} epochs, and this difference adds an artificial velocity component in the mean motion vector $\mu_{HG}$. To account for this effect, we model the offset velocity created by the difference in the star's orbital positions. Specifically, we compute the $X$ and $Y$ orbital positions at each {\it Gaia} DR2 and {\it Hipparcos} transit time with Eq.~\ref{eq:positions} (omitting observations within 1 day of each other, as in \S\ref{sec:smearing}). We then average these positions, and obtain a mean orbital position at the {\it Gaia} epoch ($\overline{X_G}$), and a mean orbital position at the {\it Hipparcos} epoch ($\overline{X_H}$). These positions are defined relative to the barycentre. The corresponding correction term is the difference between $\overline{X_G}$ and $\overline{X_H}$, divided by the time baseline between them, $\Delta{t}_{HG}$, which is taken to be 24.25 yr. This correction appears in the mean motion vector $\mu_{HG}$. For instance, at the {\it Gaia} DR2 epoch, we have
\begin{equation}
\begin{aligned}
    \mu_{HG, \delta} &= u_{\delta} + \frac{\overline{X_G} - \overline{X_H}}{\Delta{t}_{HG}}, \\
    \mu_{HG, \alpha} &= u_{\alpha} + \frac{\overline{Y_G} - \overline{Y_H}}{\Delta{t}_{HG}}.
    \label{eq:pos_corr}
\end{aligned}
\end{equation}
To first order, if the orbital phase difference is small, the mean motion vector $\mu_{HG}$ reduces to the systemic velocity $u$. The phase difference is generally larger for objects with $P > 24.25$ yr. Note that since the amplitude of the position vectors $X$ and $Y$ are proportional to $m a$ (see Eq.~\ref{eq:positions}), contributions from the inner planets to the offset factor are negligible in both $\pi$ Men and HAT-P-11.

\subsubsection{Final model of PMa}
Combining Eq.~\ref{eq:smear_corr} and Eq.~\ref{eq:pos_corr}, we can rewrite the PMa via Eq.~\ref{eq:pma_def}. At the {\it Gaia} epoch, the two-component PMa is given by
\begin{equation}
    \begin{aligned}
    \Delta\mu_{G,\delta} &= \overline{v_{G,X}} - \frac{\overline{X_G} - \overline{X_H}}{\Delta{t}_{HG}}, \\
    \Delta\mu_{G,\alpha} &= \overline{v_{G,Y}} - \frac{\overline{Y_G} - \overline{Y_H}}{\Delta{t}_{HG}}.
    \label{eq:pma_model}
    \end{aligned}
\end{equation}
Similar equations are true at the {\it Hipparcos} epoch. The systemic velocity $u$ is assumed to be linear, and is therefore cancelled out in the subtraction. A significant PMa is indicative of orbital motion caused by one or more companions. The right-hand side of Eq.~\ref{eq:pma_model} represents the model that we fit to the observed PMa data on the left-hand side of Eq.~\ref{eq:pma_model}. Our model is a function of the orbital elements, as well as the primary mass and the parallax. The observed data, taken from B18, are given in Table~\ref{tab:pma_data}. The $\Delta \mu_\alpha$ components have the $\cos{\delta}$ factor included. When combined with a radial velocity measurement, we can solve for all the orbital elements.

Several recent studies have carried out orbit fits with such data. K19 used their catalogue to recover many known companions, and place constraints (in terms of mass and semi-major axis, which are degenerate in the amplitude $A$, see Eq.~\ref{eq:orb_v}) on new candidate companions. Furthermore, proper motion anomalies have been modelled in combination with radial velocity and relative astrometry data from imaging to measure dynamical masses of known companions \citep[e.g.][]{dupuy_model_2019, brandt_precise_2019, brandt_dynamical_2019}. Independently, \citet{feng_detection_2019} combined RV data with PMa data to boost the detection significance of a newly discovered CJ around $\epsilon$ Indi A. Using measured orbital parameters from RV, \citet{kervella_orbital_2020} placed a rough constraint on the inclination of the planet candidate Proxima Centauri c.

\begin{table}
\setlength{\tabcolsep}{4pt}
\centering
\caption{Proper motion anomalies in declination and right ascension for $\pi$ Men, the validation target HD 4747, and HAT-P-11 from B18. The $\Delta \mu_\alpha$ components have the $\cos{\delta}$ factor included. $\sigma[\Delta\mu]$ represent the uncertainties. We assume that the uncertainties on the proper motions and $\mu_{HG}$ are independent and add them in quadrature to calculate these uncertainties.}
\label{tab:pma_data}
\begin{tabular}{lcccccc}
\hline
Name & Data & $\Delta\mu_{\delta}$ & $\sigma[\Delta\mu_{\delta}]$ & $\Delta\mu_{\alpha}$ & $\sigma[\Delta\mu_{\alpha}]$ & S/N\\
& epoch & \multicolumn{2}{c}{\masyr} & \multicolumn{2}{c}{\masyr} &\\
\hline
$\pi$ Men & {\it Gaia} & -0.739 & 0.263 & 0.591 & 0.246 & 3.7 \\
$\pi$ Men & {\it Hipparcos} & 0.404 & 0.445 & 0.884 & 0.398 & 2.4 \\
HD 4747 & {\it Gaia} & -1.714 & 0.588 & 3.066 & 0.552 & 6.3 \\
HD 4747 & {\it Hipparcos} & -5.118 & 0.588 & 1.316 & 0.737 & 8.8 \\
HAT-P-11 & {\it Gaia} & 0.056 & 0.091 & -0.250 & 0.084 & 3.0 \\
HAT-P-11 & {\it Hipparcos} & -1.581 & 0.853 & 0.331 & 0.905 & 1.9 \\
\end{tabular}
\end{table}

We briefly compare the catalogues from B18 and K19. Both take into account the projection effect of the stellar trajectory using the radial velocity and parallax, which is important for nearby stars like $\pi$ Men ($\approx 18.3$ pc). Unlike K19, however, B18 uses a mixture of the two {\it Hipparcos} reductions, and shows this is statistically preferred compared to a single {\it Hipparcos} reduction. We find that for $\pi$ Men, the uncertainties on the PMa from B18 are about 70 per cent higher than those quoted by K19. This is because B18 also fits for a local cross-calibration offset between the {\it Gaia} DR2 and composite {\it Hipparcos} astrometry to bring them into a common reference frame, which necessitated error inflation terms. We adopt the data from B18\footnote{Specifically, we use data published in the erratum to B18 \citep{brandt_erratum_2019}, which corrects an error in the originally published data.} instead of K19 for our analysis (listed in Table~\ref{tab:pma_data}), since the uncertainties from this catalogue are shown to be statistically well-behaved (see fig. 9 of B18), and this catalogue has been demonstrated to provide accurate orbit fits for several known companions \citep{brandt_precise_2019}. We note that the B18 catalogue does not provide values for $\Delta\mu_G$ or $\Delta\mu_H$ directly, but these can be easily derived from the calibrated proper motions $\mu_G$, $\mu_H$, and $\mu_{HG}$ in the catalogue (using Eq.~\ref{eq:pma_def}). For the uncertainty on $\Delta\mu_G$ and $\Delta\mu_H$, we follow B18 in assuming that the errors on the proper motions and $\mu_{HG}$ are independent and add them in quadrature.

\section{Measurements of Orbital Inclination} \label{sec:orbit_fits}
In this section, we perform orbit fits for three systems using the method described in \S\ref{sec:basic_method}. The first system is HD 4747, which is a validation case. The two science targets are $\pi$ Men and HAT-P-11. We focus our paper on $\pi$ Men as its tangential velocity has a higher signal-to-noise ratio (S/N), but also analyse HAT-P-11 as a comparison target since we find it to be dynamically similar to $\pi$ Men.

\subsection{Simple calculation} \label{sec:simple_calc}
We first demonstrate a simple calculation that provides a rough estimate of $I$ and $m$. In this simple calculation, we ignore the two corrections described in \S\ref{sec:smearing} and \S\ref{sec:phase_diff}, and treat the proper motion anomaly as equal to the stellar orbital velocity at the exact epochs of $t=2015.5$ for {\it Gaia} DR2 and $t=1991.25$ for {\it Hipparcos}. Therefore, to first order, the PMa is given by
\begin{equation}
\begin{aligned}
    \Delta\mu_{G,\delta} &= v_{G,X} (t=2015.5), \\
    \Delta\mu_{G,\alpha} &= v_{G,Y} (t=2015.5),
\end{aligned}
\end{equation}
at the {\it Gaia} DR2 epoch. The unit conversion between the velocities in \ms, and the PMa in \masyr is given by
\begin{equation}
     \ms = \masyr \times \frac{4740.47}{\pi},
\end{equation}
where $\pi$ is the parallax in mas (implicitly over 1 yr) and the constant term is equal to 1 au (in m) / 1 yr (in s).

From orbital energy conservation, one can derive the vis-viva equation
\begin{equation}\label{eq:v_rel}
    v_{\rm{rel}}^2 = G(M_\star+m)\left(\frac{2}{r} - \frac{1}{a}\right),
\end{equation}
where $r$ is the orbital distance between the two bodies (star and outer planet in this case), which is given by Eq.~\ref{eq:r_star} but without the leading mass ratio. $v_{\rm{rel}}$ is the relative orbital speed, and is related to the stellar orbital speed by
\begin{equation}
    v_{\star} = \frac{m}{M_\star+m} v_{\rm{rel}}.
    \label{eq:mass_ratio}
\end{equation}
We can compute $v_\star$ at $t=2015.5$ or $t=1991.25$ because we have measurements for all three components of $v_\star$ at those times
\begin{equation}
    v_{\star}^2 = v_X^2 + v_Y^2 + v_Z^2 = \Delta\mu_{\delta}^2 + \Delta\mu_{\alpha}^2 + RV^2,
    \label{eq:u_norm}
\end{equation}
where $RV$ is the radial velocity and $\Delta\mu$ is the PMa, which has two components. The RV is not necessarily known at the {\it Gaia} or {\it Hipparcos} epochs, but can be computed given a periodic RV time series. Substituting Eq.~\ref{eq:v_rel} into Eq.~\ref{eq:mass_ratio}, and using the approximation $\sqrt{M_\star+m} \approx \sqrt{M_\star}$, we can rearrange to get
\begin{equation}\label{eq:true_mass}
    m \approx v_\star \sqrt{\frac{M_\star}{G(\frac{2}{r} - \frac{1}{a})}}.
\end{equation}
We apply this calculation to our three targets, using the epoch with a higher S/N for each target (see Table~\ref{tab:pma_data}; {\it Gaia} for $\pi$ Men and HAT-P-11, and {\it Hipparcos} for HD 4747). We use orbital parameters and the stellar masses given in \citet{peretti_orbital_2019}, \citet{huang_tess_2018}, and \citet{yee_hat-p-11_2018} for HD 4747, $\pi$ Men, and HAT-P-11 respectively (listed in Table~\ref{tab:orbit_params}). This gives $m \approx 74.5~\Mj$ and $I_{B} \approx 43\degr$ for HD 4747 B, $m \approx 12.8~\Mj$ and $I_{b} \approx 52\degr$ for $\pi$ Men b, and $m \approx 2.7~\Mj$ and $I_{c} \approx 37\degr$ for HAT-P-11 c. Our rough estimate of the inclination and mass of HD 4747 is close to literature values in \citet{peretti_orbital_2019} and \citet{brandt_precise_2019}.

Given the mostly edge-on orbit of the transiting inner planets $\pi$ Men c ($I_{c} \approx 87.5\degr$) \citep{huang_tess_2018} and HAT-P-11 b ($I_{b} \approx 89.0\degr$) \citep{huber_discovery_2017}, our simple calculation already suggests that the $\pi$ Men and HAT-P-11 systems might be highly misaligned. Although the exact mutual inclination depends on the relative positions of the ascending nodes (see \S\ref{sec:pimen_fits}), $\pi$ Men and HAT-P-11 could be in contrast to the flatter mutual inclination distribution derived for typical {\it Kepler} planets \citep[e.g.][]{fabrycky_architecture_2014, dai_larger_2018-1}. We perform a more rigorous calculation in the next section, and try to understand the implications of these results in the rest of this paper.

\subsection{More detailed calculation} \label{sec:complex_fits}
In this section, we use the complete model in Eq.~\ref{eq:pma_model} to fit the PMa data, and fit for the PMa from both epochs, $\Delta \mu_G$ and $\Delta \mu_H$. These amount to two separate data points for the tangential velocity. In addition, we perform a joint fit of the PMa data with the RV time series for each of our targets.

We run our fits using the Parallel-Tempered Monte Carlo Markov Chain (PTMCMC) as implemented in \texttt{emcee v2.2.0} \citep{foreman-mackey_emcee:_2013}, which uses the algorithm described in \citet{earl_parallel_2005}. PTMCMC runs multiple MCMCs at different `temperatures,' corresponding to different modified likelihoods for the posterior. During the fit, the different MCMCs can swap information to more efficiently sample different regions of the parameter space. We use 30 different temperatures to sample the parameter space, and our results are taken from the `coldest' chain, which corresponds to the original, unmodified likelihood function. For our data, we find that PTMCMC is much more robust than a regular MCMC because our PMa and RV data can have very different S/N levels, and this results in the fits being trapped in local minima. By sampling more of the parameter space, PTMCMC avoids such traps \citep{earl_parallel_2005}. For each of the 30 temperatures, we use 50 walkers to sample the model over \SI{3e5} steps for HD 4747, and \SI{8e5} steps for $\pi$ Men and HAT-P-11, since the latter two targets have four more parameters (see below).

There are nine physically meaningful parameters in our models. The first two are the primary mass ($M_\star$) and the companion mass ($m$). We include the primary mass with a Gaussian prior based on literature values, as tabulated in Table~\ref{tab:orbit_params}. Six other parameters are needed to define the orbit: the orbital period ($P$), the longitude of ascending node ($\Omega$), the inclination ($I$), the time of periastron ($T_{p}$), and the eccentricity ($e$) and argument of periastron ($\omega_\star$) fitted as $\sqrt{e}\cos{\omega_\star}$ and $\sqrt{e}\sin{\omega_\star}$. Lastly, we include the parallax ($\pi$) in our fits with a Gaussian prior based on the {\it Gaia} DR2 astrometry \citep{gaia_collaboration_2018}. Except for the parallax and primary mass, we impose uniform priors on the other fitted parameters, using a prior in $\cos{I}$ for inclination.

Besides the physically meaningful parameters, we fit two additional parameters for each set of RV data, the instrument jitter and zero-point offset. A different jitter and offset is assigned to each different instrument. The RV data for HD 4747 comes from one instrument, so there is a total of 11 parameters for HD 4747, while $\pi$ Men has 15 parameters because its RV come from three different sources. For $\pi$ Men, we ignore the inner planet when fitting the RV as its velocity amplitude is only $\sim1$ per cent as large as that from the outer planet; its signal would simply be absorbed in the jitter. For HAT-P-11, the RV data comes from one instrument, but it is necessary to fit three additional parameters for the inner planet and one for stellar activity (see \S\ref{sec:hatp_I} for details). To combine the RV and PMa data, we adopt the same log likelihood function as \citet{brandt_dynamical_2019}:
\begin{equation}
    ln\curL = -\frac{1}{2}(\chi^2_{\Delta \mu}+ \chi^2_{RV}),
\label{eq:loglike}
\end{equation}
with
\begin{equation}
    \chi^2_{\Delta \mu} = \sum_{j}^{G, H} \left[\left(\frac{\curM[\Delta \mu_{j,\alpha}] - \Delta \mu_{j,\alpha}}{\sigma[\Delta \mu_{j,\alpha}]}\right)^{2} + \left(\frac{\curM[\Delta \mu_{j,\delta}] - \Delta \mu_{j,\delta}}{\sigma[\Delta \mu_{j,\delta}]}\right)^{2}\right],
\label{eq:chi2_pma}
\end{equation}
as the PMa part. $\alpha$ and $\delta$ denote the two directions. The sum is taken over $j$, which represents the two epochs $G$ and $H$. $\curM[\Delta \mu]$ is the model given in Eq.~\ref{eq:pma_model}, and $\Delta \mu$ and $\sigma[\Delta \mu]$ are the PMa data and uncertainties listed in Table~\ref{tab:pma_data}. The RV $\chi^2$ (for one instrument) is given by \citep[e.g.][]{howard_nasa-uc-uh_2014}
\begin{equation}
    \chi^2_{RV} = \sum_{i}\left[{\frac{(\curM[{\rm RV}_i] - {\rm RV}_i + {\rm RV}_{\rm{offset}})^2}{\sigma_i^2 + \sigma_{\rm{jit}}^2} + \ln{2\pi (\sigma_i^2 + \sigma_{\rm{jit}}^2)}}\right],
\label{eq:chi2_rv}
\end{equation}
where $\curM[{\rm RV}_i]$, ${\rm RV}_i$, and $\sigma_i$ are the model, data, and uncertainties of the radial velocity at time $i$, $\sigma_{\rm{jit}}$ is the jitter term, and $\rm{RV}_{\rm{offset}}$ is the offset term. The RV model is given by the third component of Eq.~\ref{eq:orb_v}, noting that our $+Z$ axis points toward the observer so the sign of $v_Z$ needs to be flipped to follow the convention that a positive RV corresponds to a redshift. $\rm{RV}_{\rm{offset}}$ is equivalent to the systemic velocity plus any instrumental offset. In cases where the RV data come from multiple instruments, independent $\chi^2_{RV}$ terms from different instruments are added together.

In the next three subsections, we describe our fits and measurements of HD 4747, $\pi$ Men, and HAT-P-11. In Table~\ref{tab:orbit_params}, we list all previous measurements on our targets as well as our new measurements  of $I$, $\Omega$, and $m$ in the third column.

\begin{table}
\footnotesize
\centering
\setlength{\tabcolsep}{2pt}
\caption{First two columns: Previously measured orbital parameters for $\pi$ Men, HAT-P-11, and the validation target HD 4747. Third column: new measurements from this paper. Previous values for $\pi$ Men come from \citet{huang_tess_2018}. For HAT-P-11, most measurements come from \citet{yee_hat-p-11_2018}, except for $P_b$ and $I_b$ from \citet{huber_discovery_2017}. To avoid confusion, we indicate the inner and outer planets in these systems. For HD 4747, we list complete orbital parameters from both \citet{peretti_orbital_2019} (P19) and \citet{brandt_precise_2019} (B19). For all targets, we list the primary masses used in the previous studies and {\it Gaia} DR2 parallax measurements, both of which we adopt as Gaussian priors in our fits. In the case of \citet{brandt_precise_2019}, who included imaging data in their fits, the host star mass is dynamically derived.}
\label{tab:orbit_params}
\begin{tabular}{lccc}
\hline\noalign{\vskip 2pt}
$\pi$ Men & c (in) & b (out) & b (out, this paper) \\
\hline\noalign{\vskip 2pt}

$I$ ($\degr$) & $87.46\pm0.08$ & ... & $51.2^{+14.1}_{-9.8}$ \\[3pt]
$\Omega$ ($\degr$) & ... & ... & $105.8^{+15.1}_{-14.3}$ \\[2pt]
$\msini$ ($\Mj$) & ... & $10.02\pm0.15$ & ... \\
$m$ ($\Me$, $\Mj$) & $4.82\pm0.85$ & ... & $12.9^{+2.3}_{-1.9}$ \\[2pt]
$\omega_\star$ ($\degr$) & ... & $330.61\pm0.3$ & $151.7\pm0.9^a$ \\
$P$ (d) & $6.268\pm0.0005$ & $2093.07\pm1.73$ & $2090.3\pm2.6$\\
$T_p$ (BJD-2450000) & ... & $6317.4\pm3.0$ & $6306.5\pm7.7$ \\
$e$ & $0-0.3^b$ & $0.637\pm0.002$ & $0.644\pm0.003$ \\
$M_\star (\Msun)$ & \multicolumn{3}{c}{1.094$\pm$0.039} \\
$\pi$ (mas) & \multicolumn{3}{c}{54.705$\pm$0.067} \\

\hline\noalign{\vskip 2pt}
HAT-P-11 & b (in) & c (out) & c (out, this paper) \\
\hline\noalign{\vskip 2pt}
$I$ ($\degr$) & $88.99\pm0.15$ & ... & $135.7^{+12.1^c}_{-21.4}$ \\[3pt]
$\Omega$ ($\degr$) & ... & ... & $54.4^{+28.2^c}_{-23.0}$ \\
$\msini$ ($\Me$, $\Mj$) & $23.4\pm1.5$ & $1.6\pm0.1$ & ... \\
$m$ ($\Mj$) & ... & ... & $2.3^{+0.7}_{-0.5}$$^c$ \\
$\omega_\star$ ($\degr$) & $19\pm15$ & $143.7\pm4.9$ & $323.9\pm4.1^a$ \\
$P$ (d) & $4.8878$ & $3407^{+360}_{-190}$ & $3397^{+71}_{-64}$ \\
$T_p$ (BJD-2450000) & $4959.6\pm0.2$ & $6862\pm23$ & $6865\pm19$ \\
$e$ & $0.218\pm0.033$ & $0.601\pm0.032$ & $0.604\pm0.03$ \\
$M_\star (\Msun)$ & \multicolumn{3}{c}{0.809$\pm$0.025} \\
$\pi$ (mas) & \multicolumn{3}{c}{26.480$\pm$0.033} \\

\hline\noalign{\vskip 2pt}
HD 4747 & B (P19) & B (B19) & B (this paper) \\
\hline\noalign{\vskip 2pt}
$I$ ($\degr$) & $46.3\pm1.1$ & $49.5\pm2.3$ & $50.0\pm3.5$ \\
$\Omega$ ($\degr$) & $89.9\pm1.4$ & $91.5\pm1.8$ & $98.7\pm6.3$ \\
$m$ ($\Mj$) & $70.2\pm1.6$ & $66.3^{+2.5}_{-3.0}$ & $66.6\pm3.5$ \\
$\omega_\star$ ($\degr$) & $86.9\pm0.47$ & $87.0\pm0.5$ & $87.7 \pm0.5$ \\
$P$ (yr) & $33.1\pm0.7$ & $34.0^{+0.9}_{-1.0}$ & $33.7\pm0.9$\\
$T_p$ (BJD-2450000) & $473.9\pm5.2$ & $482.0\pm350$ & $481.0\pm4.2$ \\
$e$ & $0.732\pm0.002$ & $0.735\pm0.003$ & $0.734\pm0.003$ \\
$M_\star (\Msun)$ & $0.856\pm0.014^d$ & $0.82\pm0.075$ \\
$\pi$ (mas) & \multicolumn{3}{c}{53.184$\pm$0.126} \\

\hline
\multicolumn{4}{p{0.9\linewidth}}{$^a$ Our values for $\omega_\star$ are off by $180\degr$ compared to literature values because in our coordinate system, $+Z$ points toward the observer, which is the opposite of the radial velocity convention used in those papers.}\\
\multicolumn{4}{p{0.9\linewidth}}{$^b$ The eccentricity of $\pi$ Men c is not exactly measured, but 0.3 is the 1$\sigma$ upper limit quoted in \citet{huang_tess_2018}.}\\
\multicolumn{4}{p{0.9\linewidth}}{$^c$ We quote values corresponding to $I>90\degr$ as the $I$ and $\Omega$ distributions are bi-modal for HAT-P-11 c (see Fig.~\ref{fig:hatp_3params_Imut}).}\\
\multicolumn{4}{p{0.9\linewidth}}{$^d$ We use the stellar mass from \citet{peretti_orbital_2019} for our fits.}\\
\end{tabular}
\end{table}

\subsection{Validation with HD 4747} \label{sec:validation}
We validate our method by applying it to the brown dwarf companion HD 4747 B. \citet{peretti_orbital_2019} fully determined the orbit of the companion with a joint fit of imaging and RV data. \citet{brandt_precise_2019} also characterized the orbit of HD 4747 B by adding PMa data to the imaging and RV data, which allows a direct comparison between our model and their model (with the difference that we do not include imaging data). We use RV data for HD 4747 from Keck/HIRES \citep{vogt_hires_1994}. The data was originally made public by \citet{butler_lces_2017}, and later corrected for systematics by \citet{tal-or_correcting_2019}. We use the latter source.

Our results for $I$, $\Omega$, and $m$ are listed in the third column of Table~\ref{tab:orbit_params}, and can be compared with \citet{peretti_orbital_2019} values in the first column, and \citet{brandt_precise_2019} values in the second column. In Table~\ref{tab:orbit_params}, we also list our other measured parameters, which are all consistent within $\lesssim 1\sigma$ with the respective literature values from both sources. 

We note that \citet{brandt_precise_2019} uses the same coordinate system as we do but lists $\omega$ for the companion, instead of $\omega_\star$ (T. Brandt, private communication), and therefore we subtract $180\degr$ from their value to compare in Table~\ref{tab:orbit_params}. We do the same for the \citet{peretti_orbital_2019} value. We measure $I_{B}=50.0\pm3.5\degr$ for HD 4747 B, which is consistent within $\approx 1\sigma$ to that from \citet{peretti_orbital_2019} and \citet{brandt_precise_2019}. Our measured true mass is consistent with the two previous studies as well. Note the similarity between our measured $I_{B}$ and $m$ to those from \citet{brandt_precise_2019}, who used the same PMa data of B18, although our values have slightly larger uncertainties because we do not include imaging data. Our measurement of $\Omega=98.7\pm6.3\degr$ is off by 1.4$\sigma$ and 1.1$\sigma$ from the values in \citet{peretti_orbital_2019} and \citet{brandt_precise_2019} respectively, which is reasonable agreement given that we only fit velocities and not positions from imaging data.

\subsection{Application to {$\pi$} Men} \label{sec:pimen_fits}
Next, we apply our method to $\pi$ Men, assuming that the PMa of the host star is dominated by $\pi$ Men b. In fact, \citet{zurlo_imaging_2018} attempted to image $\pi$ Men b with SPHERE and ruled out any companions with $M>26~\Mj$ beyond 5 au. Combined with the lack of long-term trend in the RV data, it is safe to interpret any significant astrometric signal as arising from $\pi$ Men b. 

For $\pi$ Men, the RV data comes from the University College London Echelle Spectrograph \citep[UCLES;][]{diego_final_1990} and the High-Accuracy Radial-velocity Planet Searcher \citep[HARPS;][]{mayor_setting_2003}. We use UCLES data published in \citet{gandolfi_tesss_2018}, and HARPS data published in \citet{trifonov_public_2020}, which corrects for small nightly offsets of the instrument. The HARPS instrument experienced an upgrade in June 2015 which resulted in a discontinuous RV jump, so we treat pre- and post-upgrade HARPS data as coming from two different instruments.

Even though $\Delta \mu_H$ has a relatively low S/N for $\pi$ Men (S/N $\approx 2.4$, see Table~\ref{tab:pma_data}), it contains information that is otherwise unavailable from only using $\Delta \mu_G$. In particular, the knowledge of velocities at two different orbital phases allows us to ascertain the direction of the orbit, and therefore to break the prograde-retrograde degeneracy in inclination between $I$ and $180\degr-I$.

\begin{figure*}
\centering
\begin{subfigure}
  \centering
  \includegraphics[width=.4\linewidth]{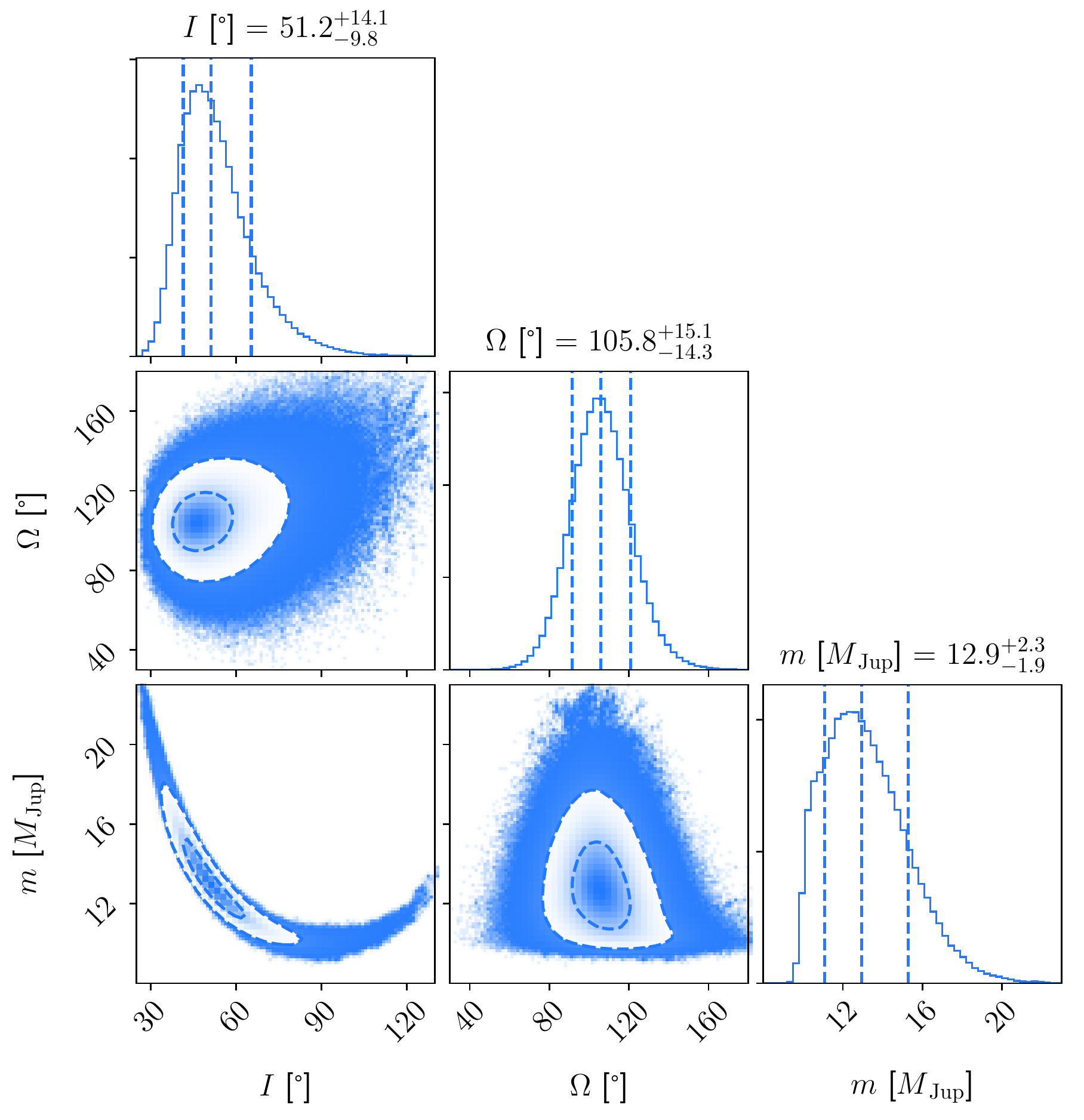}
  \hspace{10mm}
\end{subfigure}
\begin{subfigure}
  \centering
  \includegraphics[width=.4\linewidth]{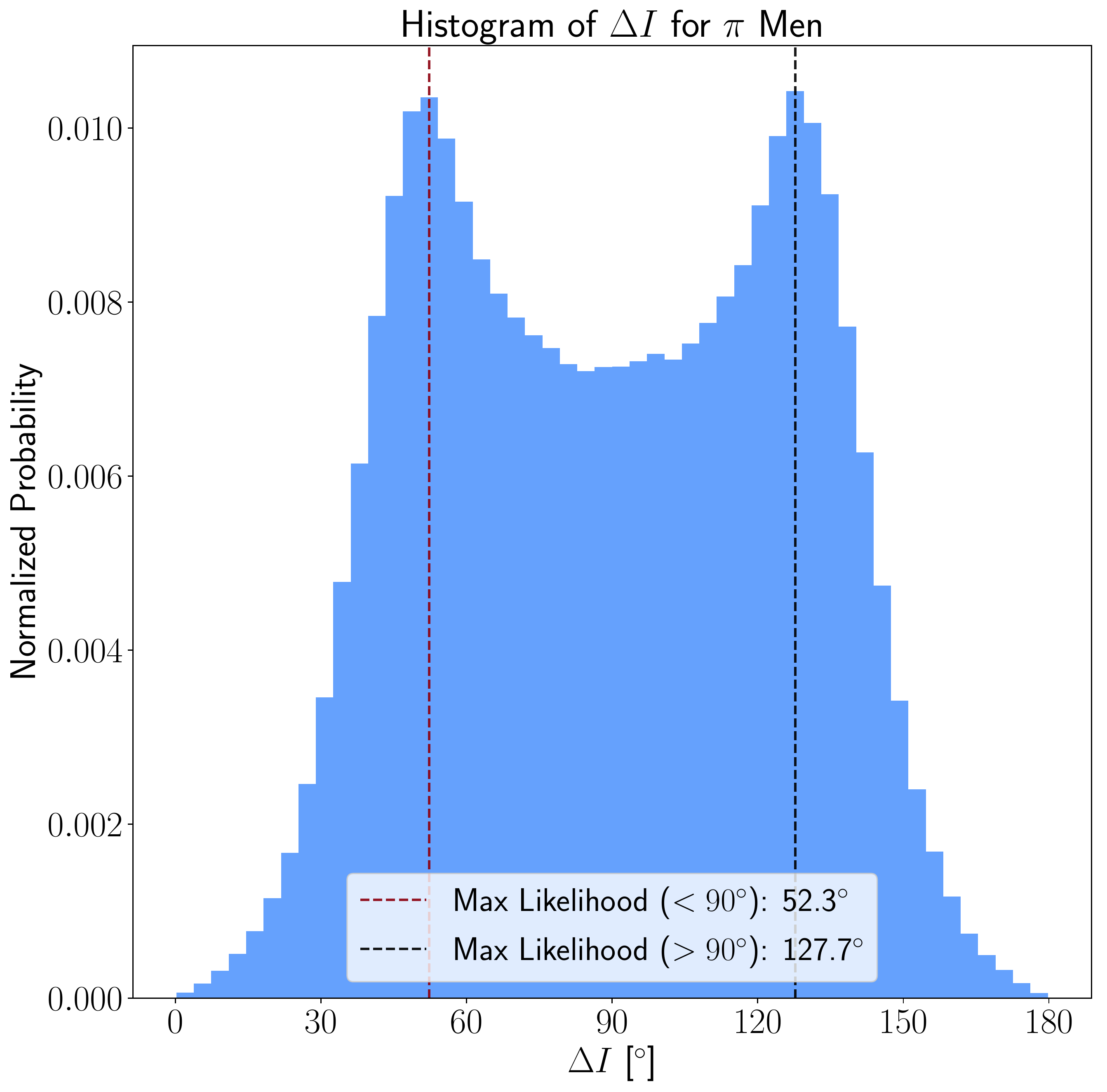}
\end{subfigure}
\caption{Target: $\pi$ Men. Left panel: Joint posterior distributions for $I$, $\Omega$, and $m$ for $\pi$ Men b. Moving outward, the dashed lines on the 2D histograms correspond to $1\sigma$ and $2\sigma$ contours. Right panel: Histogram of the mutual inclination between $\pi$ Men b and c.}
\label{fig:pimen_3params_Imut}
\end{figure*}

Our measured values are again listed in the third column of Table~\ref{tab:orbit_params}. We find that $I_{b} = 41-65\degr$ (1$\sigma$ interval), consistent with the simple calculation in \S\ref{sec:simple_calc}, and has a best fit value of $\approx47\degr$. Our results are consistent with a much broader constraint from \citet{reffert_mass_2011}, who used {\it Hipparcos} intermediate astrometric data \citep{van_leeuwen_validation_2007} and constrained the inclination to be $20-150\degr$ ($3\sigma$ interval). For the ascending node and true mass, we get $\Omega_{b}=92-121\degr$, and $m_b = 11.1-15.3$~\Mj (both 1$\sigma$ intervals). The true mass of $\pi$ Men b places it near the traditional boundary between planets and brown dwarfs, but we will still refer to it as a planet for simplicity. 

We note that our $\omega_\star$ is off by $180\degr$ compared to the \citet{huang_tess_2018} value, and this is caused by an artefact of different coordinate system definitions, as discussed in \S\ref{sec:eq_motion}.\footnote{$+Z$ points toward the observer in our coordinate system, whereas \citet{huang_tess_2018} assume that $+Z$ points away from the observer \citep[see][which the authors use]{fulton_radvel_2018}. This means that our ascending node is their descending node and vice versa, introducing a difference of $180\degr$ to $\omega_\star$.} Taking this into account, the measurements on $\omega_\star$ agree (same is true for HAT-P-11 in \S\ref{sec:hatp_I}). As shown in Table~\ref{tab:orbit_params}, all measured orbital parameters are consistent within $2\sigma$ to previous values in \citet{huang_tess_2018}. The $\approx2\sigma$ differences in $e$ and $T_p$ are likely caused by the fact that \citet{huang_tess_2018} had access to more RV data points (e.g. they used 77 UCLES RVs while we had 42). We show the complete posterior distributions for $\pi$ Men in Appendix~\ref{appendixB}, as well as distributions for the newly measured $I$, $\Omega$, and $m$ in the left panel of Fig.~\ref{fig:pimen_3params_Imut}.

Given the sky-projected inclination of the inner planet ($I_{c}$), we can put constraints on the mutual inclination $\Delta I$, which is given by
\begin{equation}
    \cos{\Delta I} = \cos{I_b}\cos{I_c} + \sin{I_b}\sin{I_c}\cos{(\Omega_b - \Omega_c)}.
    \label{eq:Imut}
\end{equation}
$\Delta I$ is dependent on the relative ascending nodes of the orbits because even if we know the transiting inner planet to be edge-on, we do not know the orientation of its orbit (i.e. the direction of its angular momentum vector) without knowledge of $\Omega_c$. In fact, the only unknown is $\Omega_c$, which can range from $0-2\pi$, meaning that $\cos(\Omega_b-\Omega_c)$ can range from -1 to 1. 

We generate a distribution for $\Delta I$ by sampling the distributions of $\Omega$ and $I$ for the two planets. For $\pi$ Men b, we use our chains from the MCMC fits. Because $I_b$ and $\Omega_b$ are not completely independent (see left panel of Fig.~\ref{fig:pimen_3params_Imut}), we randomly sample from the $I_b$ chain and take the corresponding value from the $\Omega_b$ chain, instead of randomly sampling both chains. For $\pi$ Men c, we use a bi-modal distribution for $I_c$ corresponding to the TESS measurement of $87.46\pm0.08\degr$ as well as $180\degr$ minus that to take into account the prograde-retrograde degeneracy of the inner planet. For $\Omega_c$, we uniformly sample between 0 and $2 \pi$. The right panel of Fig~\ref{fig:pimen_3params_Imut} shows the resultant distribution for $\Delta I$.  

We caution that our calculation assumes a uniform prior for $\Omega_c$, but in reality that might not be accurate as there could be an underlying prior on $\Delta I$ that disfavours values near $90\degr$, for instance on dynamical grounds. This should be taken into account when interpreting Fig~\ref{fig:pimen_3params_Imut}, as values near $\Delta I=90\degr$ might not be as likely as they appear in the figure. Overall, the distribution is symmetric about $90\degr$ due to the unknown direction of the inner orbit (i.e. retrograde or prograde with respect to the outer orbit). From the distribution, we find that $49\degr < \Delta I < 131\degr$ at the 1$\sigma$ level, $28\degr < \Delta I < 152\degr$ at the 2$\sigma$ level, and $9\degr < \Delta I < 171\degr$ at the 3$\sigma$ level. The distribution peaks near $52\degr$ and $128\degr (180\degr-52\degr)$. The large misalignment between the orbital planes of the two planets motivates us to closely examine the dynamics of the system in \S\ref{sec:analytic_dynamics}, \S\ref{sec:simulations}, and \S\ref{sec:nod_prec_effects}.

\begin{figure*}
\centering
\begin{subfigure}
  \centering
  \includegraphics[width=.4\linewidth]{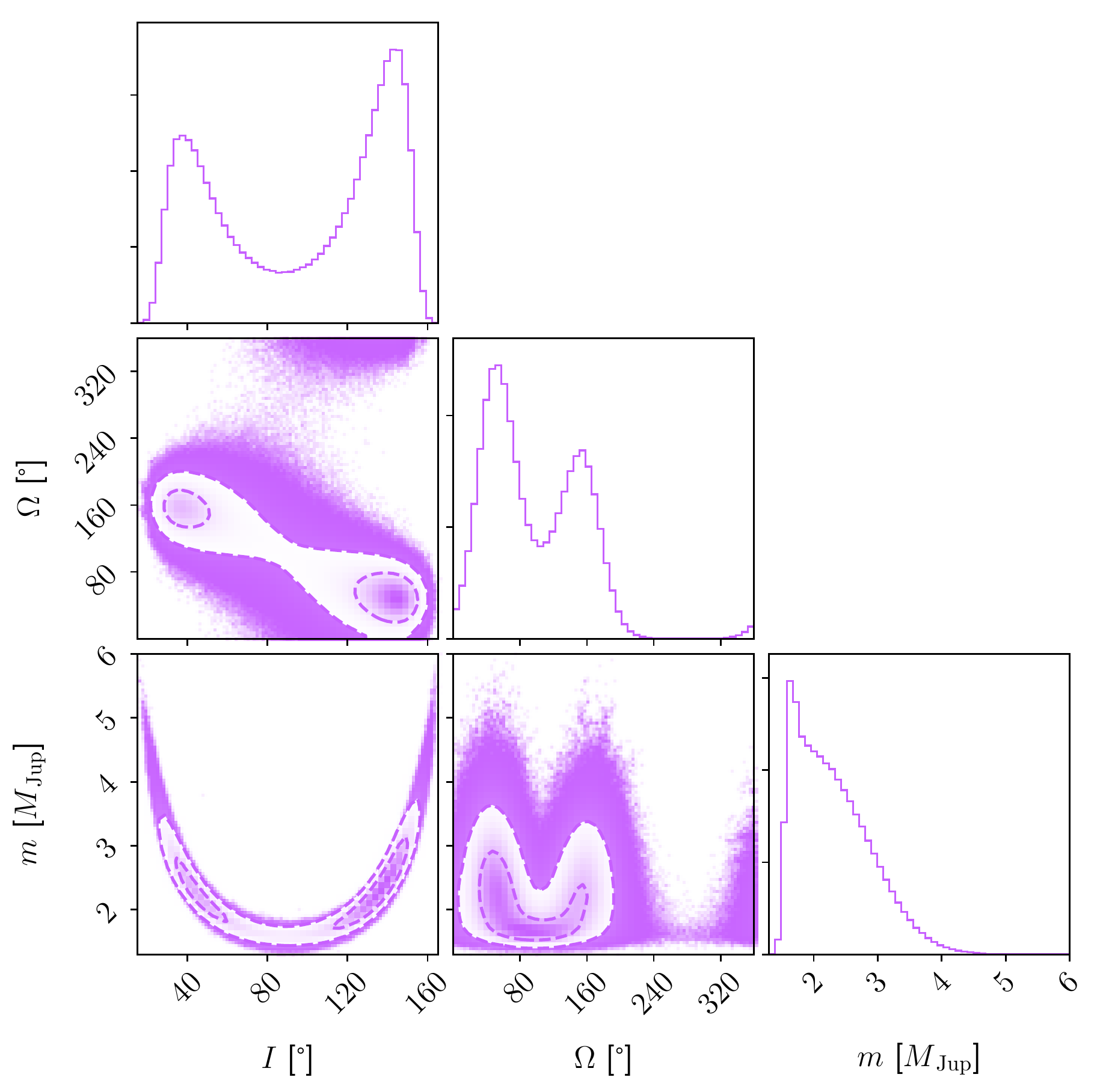}
  \hspace{10mm}
\end{subfigure}
\begin{subfigure}
  \centering
  \includegraphics[width=.4\linewidth]{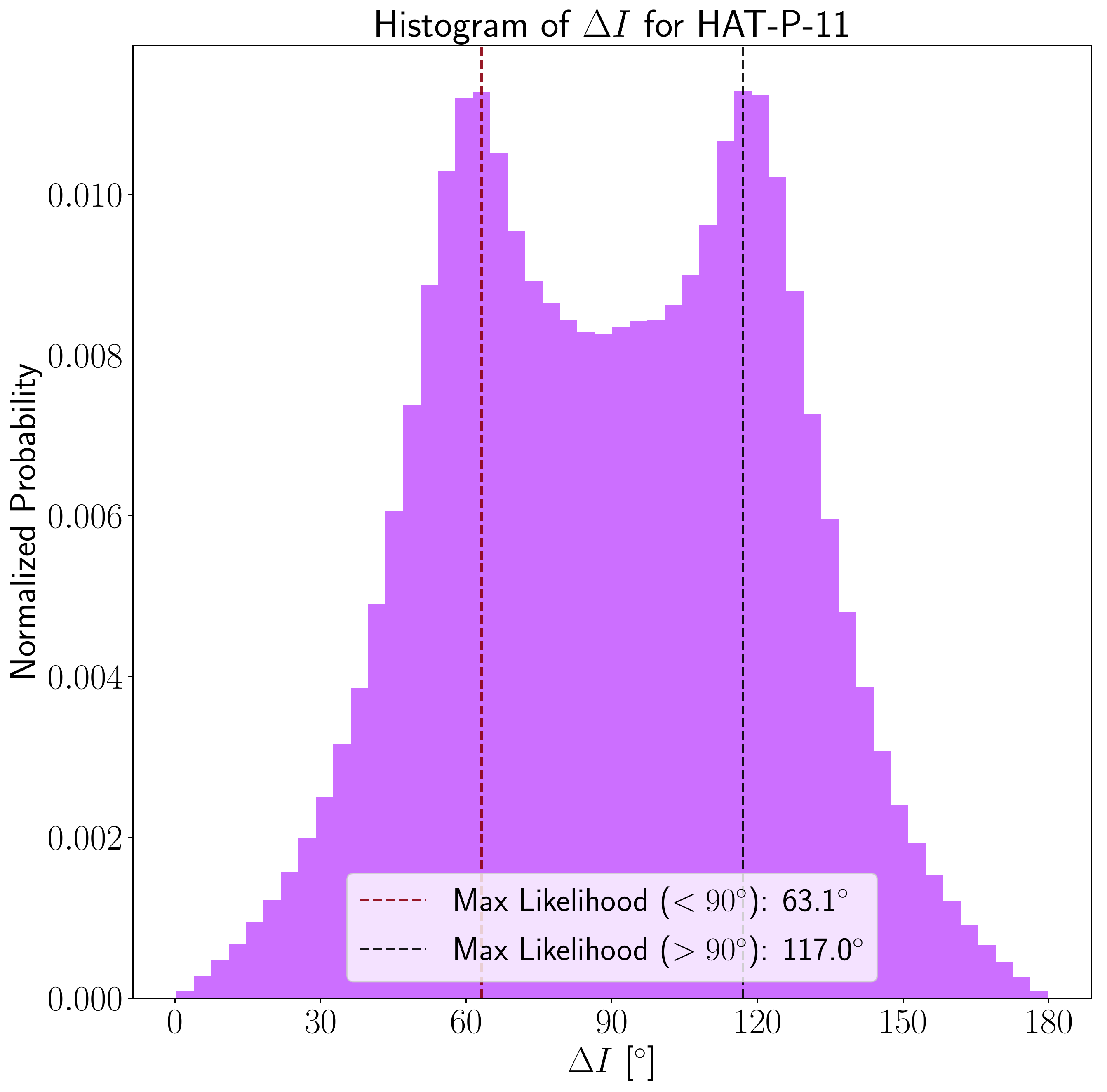}
\end{subfigure}
\caption{Target: HAT-P-11. Left panel: Joint posterior distributions for $I$, $\Omega$, and $m$ for HAT-P-11 c. Moving outward, the dashed lines on the 2D histograms correspond to $1\sigma$ and $2\sigma$ contours. Because the distributions are bi-modal, we omit the $1\sigma$ intervals in the titles. Right panel: Histogram of the mutual inclination between HAT-P-11 b and c.}
\label{fig:hatp_3params_Imut}
\end{figure*}

\subsection{Additional target: HAT-P-11}\label{sec:hatp_I}
HAT-P-11 is a system with one transiting Neptune-sized planet ($b$, $a\approx0.053$ au, $m\approx23.4~\Me$, $R\approx4.36~\Re$) first detected by \citet{bakos_hat-p-11b_2010} but with orbital elements refined by \citet{huber_discovery_2017} and one eccentric outer cold Jupiter ($c$, $a\approx4.1$ au, $\msini\approx1.6~\Mj$, $e\approx0.60$) detected in RV by \citet{yee_hat-p-11_2018}. Interestingly, the host star has a sky-projected obliquity of $\lambda \approx 100\degr$ with respect to the inner Neptune \citep{winn_oblique_2010, hirano_possible_2011}. We define the true obliquity ($\psi$) as the angle between the stellar spin axis and the orbital angular momentum axis, whereas the sky-projected obliquity ($\lambda$) is the angle between the sky-projected stellar spin axis and the sky-projected orbital axis. Since $\psi$ cannot be measured directly, $\lambda$ is used as a proxy for $\psi$. In addition, we always define the stellar obliquity with respect to the orbital axis of the inner planet in this paper.

Although HAT-P-11 does not fit perfectly in the SE+CJ framework as $\pi$ Men does, the two systems are both hierarchical ($a_{\rm{out}} \gg a_{\rm{in}}$) and similar in terms of their dynamics, and we will draw frequent comparisons between them in \S\ref{sec:analytic_dynamics}, \S\ref{sec:simulations}, and \S\ref{sec:nod_prec_effects}. In particular, \citet{yee_hat-p-11_2018} argued that nodal precession could explain the nearly pole-on orbit of HAT-P-11 b, but this explanation requires that $\Delta I \gtrsim 50\degr$ between the two planets. In this scenario, the inner orbit precesses around the angular momentum axis of the outer orbit, which approximates the invariable plane (see \S\ref{sec:Omega_precess} for details).

We find that HAT-P-11 has a proper motion anomaly with a S/N of $\approx3.0$ at the {\it Gaia} epoch, and $\approx1.9 $ at the {\it Hipparcos} epoch, weaker than the signal from $\pi$ Men (see Table~\ref{tab:pma_data}). This is expected given the smaller mass and larger semi-major axis of the outer planet in HAT-P-11 compared to that in $\pi$ Men. We use RV data published in \citet{yee_hat-p-11_2018} from Keck/HIRES, and perform orbit fits for HAT-P-11 to see if evidence for a misalignment can be found with the current data. 

For HAT-P-11, we fit a two-Keplerian model to the RV data by adding the contributions from both planets, since the RV semi-amplitudes from the two planets are comparable ($K_{\rm{in}}\approx10~\ms$ and $K_{\rm{out}}\approx30~\ms$). None the less, we ignore the inner planet when fitting the PMa data, as the contribution from the inner planet to the PMa is smeared out due to its short period (see \S\ref{sec:smearing}). Since we are interested in the outer planet (HAT-P-11 c), we follow \citet{yee_hat-p-11_2018} and fit only three parameters for the inner planet ($K$, $e$, and $\omega_\star$) and use median-values for other parameters from \citet{huber_discovery_2017}. In addition, we follow \citet{yee_hat-p-11_2018} by fitting the stellar activity as a linear trend using the $S_{\rm{HK}}$ time series published in their paper. The $S_{\rm{HK}}$ index measures the amount of emission in the Ca II H\&K lines, and is a standard tracer for stellar activity. This gives $3+1=4$ additional parameters, and 15 total parameters for HAT-P-11.

Unlike the case for $\pi$ Men, our fits of HAT-P-11 result in bi-modal distributions for $I_c$ and $\Omega_c$, as shown in the left panel of Fig.~\ref{fig:hatp_3params_Imut}. The reason why the prograde-retrograde degeneracy remains for HAT-P-11 c but not for $\pi$ Men b might be caused by the higher absolute uncertainty of its $\Delta \mu_H$ vector. As shown in Table~\ref{tab:pma_data}, the $\Delta \mu_H$ measurement errors for HAT-P-11 (row 6) are about two times larger than the $\pi$ Men $\Delta \mu_H$ errors (row 2). The $I_c$ distribution in the left panel of Fig.~\ref{fig:hatp_3params_Imut} is roughly symmetric about $90\degr$, although $I>90\degr$ solutions are slightly preferred. If we only consider solutions with $I>90\degr$, we get $114\degr < I_c < 148\degr$ ($1\sigma$). Alternatively, for solutions with $I<90\degr$, we get $33\degr < I_c < 69\degr$ ($1\sigma$). For clarity, we only quote solutions with $I > 90\degr$ in Table~\ref{tab:orbit_params}. Our measurements for all other parameters are consistent with those quoted in \citet{yee_hat-p-11_2018} to within $1\sigma$, taking into account the $180\degr$ difference in $\omega_\star$ as described in the previous subsection.

We repeat the sampling process described in the previous subsection to generate a distribution for $\Delta I$, using values of $I_{b}$ measured by \citet{huber_discovery_2017}, and drawing $\Omega_{b}$ uniformly from $0-2\pi$. As before, the $\Delta I$ distribution is bi-modal (see Fig.~\ref{fig:hatp_3params_Imut}). We find that $54\degr < \Delta I < 126\degr$ at a $1\sigma$ level in HAT-P-11, implying that there is evidence for a $\gtrsim 50\degr$ misalignment between HAT-P-11 b and c at the $\gtrsim 1\sigma$ level. This marginally supports nodal precession as the cause of the high obliquity in HAT-P-11, although a more robust conclusion would await the release of better data such as {\it Gaia} epoch astrometry. We further explore the effects of nodal precession in \S\ref{sec:Omega_precess} and \S\ref{sec:nod_prec_effects}, as we find that the same process is important in $\pi$ Men as well. 

\section{Dynamics of the $\pi$ Men and HAT-P-11 systems: analytic overview} \label{sec:analytic_dynamics}
In this section, we give an overview of the relevant dynamics in $\pi$ Men and HAT-P-11 under the secular approximation, where energy (semi-major axis) is conserved and the orbits exchange angular momentum on time scales much longer than the orbital periods. In \S\ref{sec:tidal_diss}, we also consider whether energy loss due to non-conservative tidal dissipation is important in our systems. An analytic study of the dynamics in HAT-P-11 has been performed by \citet{yee_hat-p-11_2018}, so we focus our discussion on $\pi$ Men in this section and use HAT-P-11 for comparison, summarising its relevant time scales in \S\ref{sec:hatp_analytic}.

When discussing the dynamics, it is potentially confusing that planet c is the inner planet in $\pi$ Men while planet b is the inner planet in HAT-P-11. Therefore, for clarity, we shall henceforth denote orbital elements of the inner and outer orbits with subscripts 1 and 2, respectively (e.g. $e_1$, $a_1$ are the eccentricity and semi-major axis of the inner orbit). Similarly, we denote the masses of the inner and outer planets as $m_1$ and $m_2$, while the stellar mass is $M_\star$.

The $\pi$ Men and HAT-P-11 systems are both hierarchical triple systems, where the ratio of the inner and outer semi-major axes is a small parameter. For the $\pi$ Men planets, $a_1 / a_2 \approx 0.021$,\footnote{We note that there is a typo in \citet{huang_tess_2018} for $a_2$. The listed value is 3.10 au, while it should be 3.31 au (C. Huang, private communication).} and for HAT-P-11, $a_1 / a_2 \approx 0.013$. A hierarchical triple system can be treated as two binaries: an inner binary and an outer binary between the tertiary object and the centre of mass of the inner binary. For our two systems, the mutual inclinations are large between the inner and outer binaries (see right panels of Fig.~\ref{fig:pimen_3params_Imut} and Fig.~\ref{fig:hatp_3params_Imut}).

We use the invariable plane as the reference frame in which to understand the dynamics. The invariable plane is the plane perpendicular to the direction of total angular momentum of the system, which is well-approximated by the sum of the orbital angular momentum of the two planets, $L_1 + L_2$. Conveniently, in the invariable plane, $\Delta I = I_1 + I_2$, where $I_1$ and $I_2$ are the inclinations of the inner and outer orbits with respect to the invariable plane. In $\pi$ Men and HAT-P-11, the outer planets dominate the total angular momentum, as $L$ is proportional to $ma^{1/2}$. In the next three subsections (\S\ref{sec:t_kl}-\S\ref{sec:tidal_diss}), we calculate relevant time scales for $\pi$ Men.

\subsection{Secular perturbations from the outer CJ}\label{sec:t_kl}
Due to secular perturbations from an outer companion, the inner orbit can undergo oscillatory changes in eccentricity and inclination on time scales much longer than the orbital periods. At high mutual inclinations ($39.2\degr \lesssim \Delta I \lesssim 140.8\degr$), \citet{kozai_secular_1962} and \citet{lidov_evolution_1962} studied the evolution of the inner orbit for the case of a circular outer orbit and an inner test particle. The resulting behaviour, which is characterized by large amplitude oscillations in $e_1$ and $I_1$, can be described with a perturbing function truncated to quadrupole order in semi-major axis ratio, $(a_1/a_2)^2$. Since then, it has been shown that octupole terms proportional to $(a_1/a_2)^3$, which are nonzero in the case of eccentric outer companions, lead to more complex behaviour, including prograde-retrograde flips of the inner orbit even at low mutual inclinations (see e.g. \citealt{lithwick_eccentric_2011,katz_long-term_2011}, and the review by \citealt{naoz_eccentric_2016} on the so-called `Eccentric Kozai-Lidov Effect').

We limit our analytic consideration to the quadrupole order in order to simplify the equations and provide a rough estimate of the relevant time scales. In reality, $e_2 \approx 0.64$ for $\pi$ Men b, and we perform a more detailed examination of the system using $N$-body simulations in \S\ref{sec:simulations}. We also assume that $m_1=0$ in this subsection, which means that the orbital plane of the outer planet {\it is} the invariable plane and therefore $I_2 = 0$, and $\Delta I = I_1$. Following \citet{naoz_eccentric_2016}, we refer to the use of these two assumptions as the `test particle quadrupole' approximation (TPQ for short).

To understand the evolution of the system, it is convenient to use the Delaunay variables, which describe three angles and three conjugate momenta for each orbit. In particular, mutual secular perturbations induce apsidal precession involving $\omega$ and nodal precession involving $\Omega$ (the mean anomaly is eliminated in the secular approximation). The relevant equations of motion for the inner orbit are
\begin{equation}\label{eq:omega_dot}
    \dot{\omega}_1 = -\frac{\partial \curH}{\partial G_1}, ~~~~ G_1 = \sqrt{1-e_1^2},
\end{equation}
\begin{equation}\label{eq:Omega_dot}
    \begin{aligned}
    \dot{\Omega}_1 = -\frac{\partial \curH}{\partial H_1}, ~~~ H_1 &= \sqrt{1-e_1^2}\cos{I_1},
    \end{aligned}
\end{equation}
where $G_1$ is the scaled conjugate momentum of $\omega_1$, $H_1$ is the scaled conjugate momentum of $\Omega_1$, and $\curH$ is the Hamiltonian of the system. For $m_1=0$, the Hamiltonian is given by \citep[e.g.][]{naoz_eccentric_2016, yee_hat-p-11_2018}
\begin{equation} \label{eq:Hamiltonian}
    \curH = \frac{1}{16} \sqrt{\frac{G}{M_\star}} \left(\frac{a_1^{3/2}}{a_2^{3}}\right) \frac{m_2}{(1-e_2^2)^{3/2}} ~ f_{\rm{quad}},
\end{equation}
where
\begin{equation} \label{eq:f_quad_delaunay}
    f_{\rm{quad}} = \left[\frac{(5-3G_1^2)(3H_1^2-G_1^2)}{G_1^2} + \frac{15(1-G_1^2)(G_1^2-H_1^2)\cos{2\omega_1}}{G_1^2}\right]
\end{equation}
is the perturbing function truncated to quadrupole order.

Since the Hamiltonian does not depend on $\Omega_1$, the conjugate momentum $H_1 = \sqrt{1-e_1^2} \cos{I_1}$ in Eq.~\ref{eq:Omega_dot} is conserved in the TPQ approximation. Therefore, changes in eccentricity and inclination are coupled. For hierarchical systems with $39.2\degr \lesssim \Delta I \lesssim 140.8\degr$, large amplitude oscillations in $e_1$ and $I_1$ can take place \citep{kozai_secular_1962,lidov_evolution_1962}. Depending on the initial $\omega_1$, these oscillations occur as $\omega_1$ either librates or circulates \citep[e.g.][]{holman_chaotic_1997-1, katz_long-term_2011}, and are controlled by the term with $\cos{2\omega_1}$ in Eq.~\ref{eq:f_quad_delaunay}. In the TPQ approximation, the maximum eccentricity reached for an initial $e_1 = 0$ is \citep[e.g.][]{innanen_kozai_1997}
\begin{equation}\label{eq:kl_emax}
    e_{\rm{max}} = \sqrt{1-\frac{5}{3}\cos^2{\Delta I_0}},
\end{equation}
where $\Delta I_0$ is the initial mutual inclination. Note that $e_{\rm{max}}$ approaches 1 as $\Delta I_0$ approaches $90\degr$. Because $H_1$ is conserved, $I_1$ reaches a minimum value when $e_1$ is maximum and vice versa. In the TPQ approximation, the time scale for these large eccentricity excitations can be estimated by considering the time evolution of $\omega_1$ \citep[e.g.][]{holman_chaotic_1997-1, naoz_resonant_2013}
\begin{equation}\label{eq:t_kl}
    \tau_{\rm{quad}} = \frac{M_\star}{m_2} \left(\frac{a_2}{a_1}\right)^3 (1-e_2^2)^{3/2} P_1,
\end{equation}
where $P_1$ is the period of the inner orbit. For $\pi$ Men, we get $\tau_{\rm{quad}}$ $\sim$ \SI{7.5e4} yr.

In reality, there are additional effects beyond perturbations from the outer companion that can influence the inner orbit. In \S\ref{sec:conserve_srfs}, we consider the effect of short-range conservative forces. Then, in \S\ref{sec:tidal_diss}, we consider whether the non-conservative effect of tidal dissipation is important in $\pi$ Men.

\subsection{Conservative short-range forces: general relativity and tides}\label{sec:conserve_srfs}
To accurately model the $\pi$ Men system, we must take into account effects beyond Newtonian secular perturbations (set by $f_{\rm{quad}}$ in Eq.~\ref{eq:f_quad_delaunay}). Past studies have recognized the importance of apsidal precession caused by general relativity (GR) effects and tidal interactions between the host star and a short-period planet \citep[e.g.][]{ford_secular_2000, fabrycky_shrinking_2007, liu_suppression_2015}. These effects are conservative and can be incorporated as extra terms in the total Hamiltonian. If the extra precession occurs in the opposite direction to that induced by secular perturbations, and has a shorter time scale than $\tau_{\rm{quad}}$, large eccentricity excitations resulting from high $\Delta I$ can be suppressed \citep[e.g.][]{naoz_eccentric_2016}.

We compare the time scales of conservative short-range forces with $\tau_{\rm{quad}}$ (Eq.~\ref{eq:t_kl}) to predict the behaviour of the system. Under the first post-Newtonian expansion, the GR precession time scale is given by \citep[e.g.][]{fabrycky_shrinking_2007}
\begin{equation}\label{eq:t_gr}
     \tau_{\rm{GR}} = 2\pi/\dot{\omega}_{\rm{GR}} = \frac{a_1 c^2 (1-e_1^2)}{3 G (M_\star+m_1)} P_1 ,
\end{equation}
where $c$ is the speed of light. For $e_1$ = 0, we find that $\tau_{\rm{GR}}$ $\sim$ \SI{3.6e4} yr and decreases with higher $e_1$. Since $\tau_{\rm{GR}}$ is about a factor of two shorter than $\tau_{\rm{quad}}$, and GR precession takes place in the opposite direction to that induced by $f_{\rm{quad}}$, we expect GR effects to suppress eccentricity growth in $\pi$ Men, and therefore include them in our simulations in \S\ref{sec:simulations}.

Non-dissipative tidal torques on the inner planet from the central star could also induce apsidal precession that counteracts the evolution due to $f_{\rm{quad}}$. Assuming $M_\star \gg m_1$, and $e_1 \ll 1$, the time scale of tidal precession can be estimated as \citep[e.g.][]{pu_low-eccentricity_2019}
\begin{equation}
    \tau_{\rm{tides}} = 2\pi / \dot{\omega}_{\rm{tides}} = \frac{2}{15k_{2,1}} \frac{m_1}{M_\star} \left(\frac{a_1}{R_1}\right)^5 P_1,
    \label{eq:w_tides}
\end{equation}
where $k_{2,1}$ and $R_1$ are the tidal Love number and radius of the inner planet. For lack of a better estimate, we assume the value of $k_{2,1}$ = 0.3 measured for Earth \citep{wahr_body_1981}. This gives $\tau_{\rm{tides}}\sim$ \SI{3.5e7} yr, which is nearly three orders of magnitude longer than $\tau_{\rm{quad}}$. Therefore, we can safely ignore the effect of tidal precession in our simulations even if in reality $k_{2,1}$ is larger by an order of magnitude or so. Note that the dependence on $R^5$ means that super Earths experience much weaker tidal precession effects than hot Jupiters in general.

\subsection{Tidal dissipation}\label{sec:tidal_diss}
We now consider whether the non-conservative effect of tidal dissipation is important in $\pi$ Men. Tidal dissipation has been used to explain the existence of ultra-short period planets (USPs) via migration (specifically, with eccentricity tides). As a population, USPs have $P<1$ d (or $a<0.02$ au for Sun-like hosts), $R<2~\Re$, and rocky compositions \citep{winn_kepler-78_2018}. Two different migration scenarios have been proposed for USPs, and termed `high-eccentricity migration' \citep{petrovich_ultra-short-period_2019-1}, and `low-eccentricity migration' \citep{pu_low-eccentricity_2019}. Both rely on secular interactions to excite the eccentricity of the innermost planet and reduce its pericentre distance, thereby allowing increasingly efficient dissipation to sap its orbital energy and lead to circularisation and orbital decay. We find in \S\ref{sec:max_e} that $\pi$ Men c cannot reach the high eccentricities ($e\gtrsim0.8$) required in \citet{petrovich_ultra-short-period_2019-1}, and therefore focus on `low-eccentricity migration' in the following discussion.

\citet{pu_low-eccentricity_2019} showed that under the influence of two or more short-period outer planets with $e > 0.1$ and $m=3-20~\Me$, inner planets starting at $a=0.02-0.04$ au can readily evolve into USPs while maintaining low eccentricities. However, their maximum starting $a$ is limited to about 0.04 au because the tidal dissipation efficiency is strongly dependent on orbital distance (e.g. see Eq.~\ref{eq:e_tides}). $\pi$ Men c currently orbits at $\approx0.07$ au around its host star, which is estimated to be $\approx3$ Gyr old \citep{huang_tess_2018}. The fact that $\pi$ Men c has not evolved into a USP means that the orbital decay time scale (i.e. $\tau_a = a / \dot{a}$) is likely to be much longer than 3 Gyr. However, eccentricity damping (i.e. $\tau_e = e / \dot{e}$) is expected to take place much faster than orbital decay \citep{pu_low-eccentricity_2019}. Specifically, according to \citet{pu_low-eccentricity_2019}, $\tau_a \propto \tau_e / e^2$ at low $e$, so the eccentricity evolution is at least $\sim10$ times faster than that of the semi-major axis for $e<0.3$ (where 0.3 is the $1\sigma$ upper limit quoted in \citealt{huang_tess_2018}).

Adopting the model of equilibrium tides from \citet{hut_tidal_1981-1}, we can place a rough estimate on $\tau_e$ and therefore $\tau_a$, assuming low $e$ and zero obliquity \citep[e.g.][]{pu_low-eccentricity_2019}
\begin{equation}\label{eq:e_tides}
 \tau_{e} = \frac{e}{\dot{e}} = \frac{2 Q_1}{21 \pi k_{2,1}} \frac{m_1}{M_\star} \left(\frac{a_1}{R_1}\right)^5 P_1,
\end{equation}
where $Q_1$ is the tidal quality factor of the inner planet. The numerical value is not known, but note that the time scale is proportional to $Q_1/k_{2,1}$. If we assume a typical terrestrial value of $Q_1=100$ \citep{goldreich_q_1966-1}, and $k_{2,1} = 0.3$ as in \S\ref{sec:conserve_srfs}, we get $\tau_{e}\sim 0.7$ Gyr, which is lower but comparable to the system age of 3 Gyr. This value of $\tau_e$ would imply $\tau_a > 7$ Gyr, consistent with the fact that $\pi$ Men c has not evolved into a USP.

As the evolution of $a$ is likely insignificant, we now consider the evolution of $e$. If the above values for $k_{2,1}$ and $Q_1$ are roughly correct, and $\tau_e$ is less than the system age, we expect the free eccentricity of $\pi$ Men c, which is set by initial conditions, to have been removed by tidal dissipation. As the free eccentricity is removed, $e$ will tend to the forced eccentricity, which then powers
subsequent dissipation \citep{pu_low-eccentricity_2019}. On the other hand, if $Q_1/k_{2,1}$ is under-estimated by a factor of five or more, $\tau_e$ would exceed the system age, and eccentricity damping due to tides would not be important in $\pi$ Men. 

Given that $Q_1$ and $k_{2,1}$ could both be uncertain by an order of magnitude or more, it is unclear whether tides are important in the eccentricity evolution of $\pi$ Men c without, for instance, a better constraint on $e_1$. Due to this uncertainty, we choose not to include tidal dissipation in our simulations and instead focus on the evolution over Myr time scales, where secular perturbations dominate. The main results from our study of the dynamics (\S\ref{sec:nod_prec_effects}) are not strongly affected even if $\tau_e<3$ Gyr, and we shall point out caveats of not including tides where relevant. We encourage a detailed study of tides in $\pi$ Men, especially in comparison to USPs, for future work.

\subsection{Time scales for HAT-P-11}\label{sec:hatp_analytic}
As noted by \citet{yee_hat-p-11_2018}, large eccentricity excitations are also suppressed by GR precession in HAT-P-11. In fact, for HAT-P-11, $\tau_{\rm{GR}}$ is about 60 times shorter than $\tau_{\rm{quad}}$ ($\tau_{\rm{GR}} \sim$\SI{3e4} yr vs. $\tau_{\rm{quad}} \sim$\SI{1.8e6} yr).

The tidal precession and eccentricity damping time scales for HAT-P-11 b are difficult to estimate as the tidal properties of even our own Neptune are not well known \citep{lainey_quantification_2016}. For a rough estimate, we assume $k_{2,1} = 0.3$ and $Q_1=$ \SI{7e4} for the Neptune-sized planet, where the value of $Q_1$ is the lower limit found for Uranus \citep{goldreich_q_1966-1}. This yields $\tau_{\rm{tides}} \sim $ \SI{9e5} yr, which is about two times shorter than $\tau_{\rm{quad}}$ so eccentricity growth is further suppressed for HAT-P-11 b. Given the low eccentricity ($e\approx0.22$) of HAT-P-11 b, we use Eq.~\ref{eq:e_tides} to estimate $\tau_e$ (note that this ignores obliquity tides) and get $\tau_{e} \sim 15$ Gyr, longer than the system age of $\approx 7$ Gyr \citep{bakos_hat-p-11b_2010}. 

Indeed, the observed $e\approx0.22$ is relatively high for HAT-P-11 b given its location ($a\approx0.05$ au). As the forced eccentricity is expected to be negligibly small \citep{yee_hat-p-11_2018}, this means that the free eccentricity of HAT-P-11 b has not been removed over 7 Gyr, and therefore tidal dissipation (either in terms of eccentricity damping or orbital decay) is likely not important for the planet. Indeed, setting $\tau_e$ to the system age could provide a lower limit on $Q_1/k_{2,1}$ for HAT-P-11 b, although tides raised due to obliquity would need to be considered here \citep{millholland_obliquity_2019}, so one should use the full tidal equations provided in e.g. \citet{leconte_is_2010}.

\begin{figure*}
\centering
\includegraphics[width=0.8\linewidth]{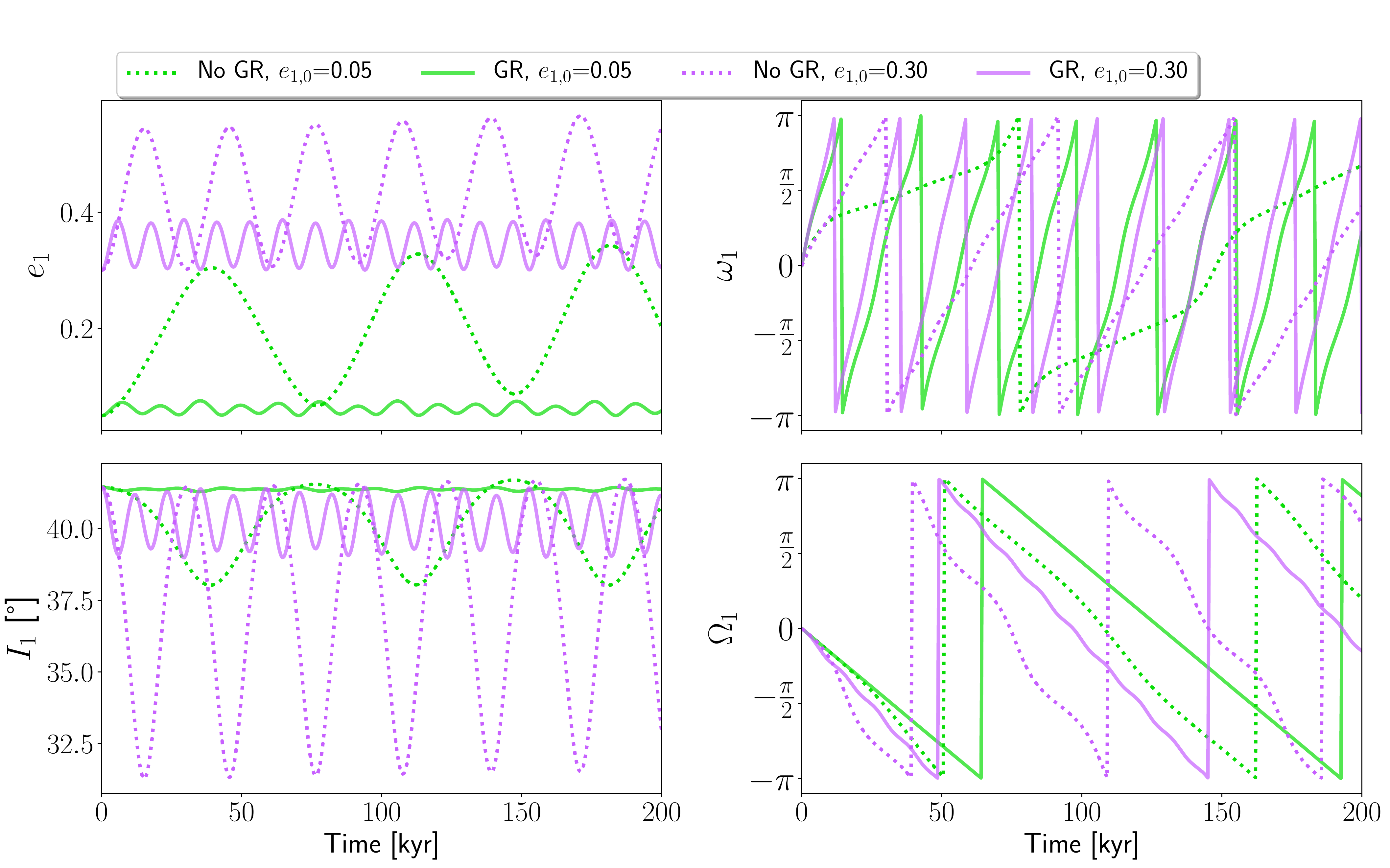}
\caption{The evolution of $\pi$ Men c with initial $\Delta I\approx41\degr$, $\omega_1=0$, $\Omega_1=0$, and two initial values of $e_1$ (green lines: 0.05; purple lines: 0.30). In top panel, $e_1$ and $\omega_1$ are plotted; in the bottom panel $I_1$ and $\Omega_1$. As a comparison, simulations without GR are shown in dashed lines.}
\label{fig:evol_all}
\end{figure*}

\section{$N$-body Simulations and Results} \label{sec:simulations}
In this section, we perform $N$-body simulations for $\pi$ Men, based on our new measurements of $I$, $\Omega$, and $m$ for the outer CJ and previous measurements on the system from \citet{huang_tess_2018}. We choose not to perform $N$-body simulations for HAT-P-11 due to the lower significance of its inclination measurement, and the fact that it has been studied in \citet{yee_hat-p-11_2018}. In this section, we first explain our simulation setup in \S\ref{sec:sim_setup}, and then describe the basic motion in \S\ref{sec:basic_motion}. We discuss the quantitative results from our simulations in \S\ref{sec:max_e} and \S\ref{sec:Omega_precess}. Then, in \S\ref{sec:nod_prec_effects}, we will discuss observational consequences of the secular interactions in our systems.

\subsection{Simulation setup}\label{sec:sim_setup}
For our $N$-body simulations, we use the \texttt{IAS}15 integrator as implemented in the \texttt{REBOUND} package \citep{rein_rebound_2012}. The \texttt{IAS}15 algorithm uses an adaptive time scale that ensures machine-level precision, and is particularly accurate in handling close encounters and high-eccentricity orbits \citep{rein_ias15_2015}. As discussed in \S\ref{sec:analytic_dynamics}, we include the effects of GR using the \texttt{REBOUNDx} package with the option \texttt{gr} \citep{tamayo_reboundx_2019}, which is valid for cases where the masses are dominated by a single central body. We start our simulations with a timestep of 2 per cent the orbital period of the inner planet, and store the orbital elements after every 500 yr of integration. Our simulations are generally run over $\sim1$ Myr.

We assign mass to the inner planet in our simulations, and thus take into account both $L_1$ and $L_2$ to calculate the exact invariable plane in this section. This means that we no longer have $I_1 = \Delta I$ as in the previous section, but instead $I_1 \approx \Delta I$ (since $L_1 \ll L_2$).

\subsubsection{Transformation between sky and invariable planes}
To run our simulations in the invariable plane, where the dynamics is more easily understood, we first transform angles in the sky plane (what is observed) to angles in the invariable plane. We include details for the plane transformation in Appendix~\ref{appendixA}. In terms of notation, we denote angles in the sky plane as $\omega_{\rm{sky}}$, $\Omega_{\rm{sky}}$, and $I_{\rm{sky}}$, and angles in the invariable plane as $\omega_{\rm{inv}}$, $\Omega_{\rm{inv}}$, and $I_{\rm{inv}}$. For example, $I_{\rm{1,sky}}$ refers to the sky-projected inclination of $\pi$ Men c (the inner planet), while $I_{\rm{1,inv}}$ refers to its inclination in the invariable plane. Parameters such as $a$, $e$, and $m$ are frame-independent.

\subsubsection{Unknown parameters and initial conditions: $\Omega_1$, $e_1$, and $\omega_1$}
The unknown or uncertain orbital elements in our system introduce uncertainties to the simulations. We take these uncertainties into account in our consideration of the dynamics.

Most importantly, the unknown ascending node of the inner planet $\Omega_{\rm{1,sky}}$ creates a range of possible $\Delta I$, as shown in Fig.~\ref{fig:pimen_3params_Imut}. Therefore, for each simulation, we sample from the distributions of the sky angles $I_{\rm{1,sky}}$, $\Omega_{\rm{1,sky}}$, $I_{\rm{2,sky}}$, and $\Omega_{\rm{2,sky}}$, as described in \S\ref{sec:pimen_fits}. For each set of sky angles, we compute $\Delta I$ using Eq.~\ref{eq:Imut}. The four sky angles also set the location of the invariable plane with respect to the sky plane (see Eq.~\ref{eq:Lxyz} and Eq.~\ref{eq:Ip_Omega_p} in Appendix~\ref{appendixA}).

The two other unknowns are $e_1$ and $\omega_{\rm{1,sky}}$ (which can be converted to $\omega_{\rm{1,inv}}$). The dynamical evolution will be different for different initial values of $e_1$ and $\omega_{\rm{1,inv}}$, since the Hamiltonian of the system contains these parameters (see Eq.~\ref{eq:Hamiltonian}). Therefore, we also sample from distributions in $e_1$ and $\omega_{\rm{1,sky}}$. We assume a flat distribution for $\omega_{\rm{1,sky}}$ from 0 to 2$\pi$, and a flat distribution for $e_1$ between 0 to 0.3, where 0.3 is the measured 1$\sigma$ upper limit from \citet{huang_tess_2018}. For the remaining orbital elements, namely $a_1$, $P_1$, $a_2$, $P_2$, $e_2$, and $\omega_{\rm{2,sky}}$, we use the median values from \citet{huang_tess_2018} listed in Table~\ref{tab:orbit_params}, since the uncertainties on these values are negligible compared to that of the unknowns.

Finally, we convert the sky angles of both orbits to angles in the invariable plane using Eq.~\ref{eq:sky_inv}, and initialize our simulations with the resultant set of orbital elements. We run a suite of 2000 separate simulations of the system in order to account for uncertainties.

\subsection{Basic picture: apsidal and nodal precession}\label{sec:basic_motion}
In our simulations, we identify two basic secular effects, apsidal precession and nodal precession, for which the equations of motion are given in Eq.~\ref{eq:omega_dot} and Eq.~\ref{eq:Omega_dot}, respectively. Under apsidal precession, $e$ oscillates between minimum and maximum values on a time scale set by $\dot{\omega}_1$. The conjugate momentum $H_1$ (Eq.~\ref{eq:Omega_dot}), which is conserved in the TPQ approximation, means that the dominant behaviour of $I$ is oscillation on the same time scale with an opposite phase. On the other hand, nodal precession causes $\Omega$ to evolve on a different (albeit similar) time scale. These evolutions occur for both planets, but at a larger amplitude for the inner planet due to its smaller angular momentum. In the following, we discuss the orbit of the inner planet.

In Fig.~\ref{fig:evol_all}, we show the evolution of the inner orbit under $\Delta I \approx 41\degr$, and two different initial values of $e_1$ (0.05 in green and 0.3 in purple). $\Omega_1$ and $\omega_1$ are both initialized to 0. Apsidal precession is shown in the top panel, where we plot $e_1$ (top left) and $\omega_1$ (top right) as functions of time. In the bottom panel, we plot the evolution of $I_1$ (bottom left) and $\Omega_1$ (bottom right) from the same simulations. In these plots, we overlay the resulting evolution when GR precession is not included for comparison purposes (shown in dashed lines).

As shown in Fig.~\ref{fig:evol_all}, both $\omega_1$ and $\Omega_1$ precess from $-\pi$ to $+\pi$. Without GR, the inner orbit shows relatively large amplitude oscillations of $e_1$ and $I_1$ that occur in opposite phases (see dashed lines in left two panels). As discussed in \S\ref{sec:conserve_srfs}, $\tau_{\rm{GR}}$ is shorter than $\tau_{\rm{quad}}$ for $\pi$ Men c, which causes the apsidal precession rate to be higher when GR is included. This can be seen by comparing the periods of the solid and dashed lines in the top left panel for example. The faster GR precession rate suppresses eccentricity growth (and consequently, large fluctuations in inclination). Indeed, Fig.~\ref{fig:evol_all} shows that the amplitude of oscillations in $e_1$ and $I_1$ are much reduced when GR is included. For example, without GR, $I_1$ can fluctuate by $\approx 10\degr$ with an initial $e_1=0.3$ (dashed purple line in bottom left panel), while $I_1$ only changes by $\approx 2\degr$ with GR (solid purple line in same panel). Similarly, $e_1$ always reaches some maximum value during the evolution, but this maximum value is lower than it would be if GR had not been included (see top left panel).

We note that Fig.~\ref{fig:evol_all} shows the evolution for less than 1 Myr, without the effect of tidal dissipation. If tidal dissipation were included, the value about which $e_1$ oscillates would slowly decrease with time so that after $\tau_e$, which we estimate to be $\sim0.7$ Gyr but is highly uncertain (\S\ref{sec:tidal_diss}), the initial conditions would be forgotten and both the purple and green lines in the top left panel would be oscillating about the same (low) value determined by the forced eccentricity. From other simulations with initial $e_1\approx0$ and $\Delta I\approx41\degr$, we find that the forced eccentricity is $\lesssim0.01$ in this case.

\subsection{Maximum eccentricity}\label{sec:max_e}
Although GR precession suppresses eccentricity growth caused by $f_{\rm{quad}}$, $e_1$ is still excited to some extent. In this section, we examine the maximum $e_1$ ($e_{\rm{max}}$) reached in our simulations, and compare the results to previous analytic estimates. 

\begin{figure}
    \centering
    \includegraphics[width=0.9\linewidth]{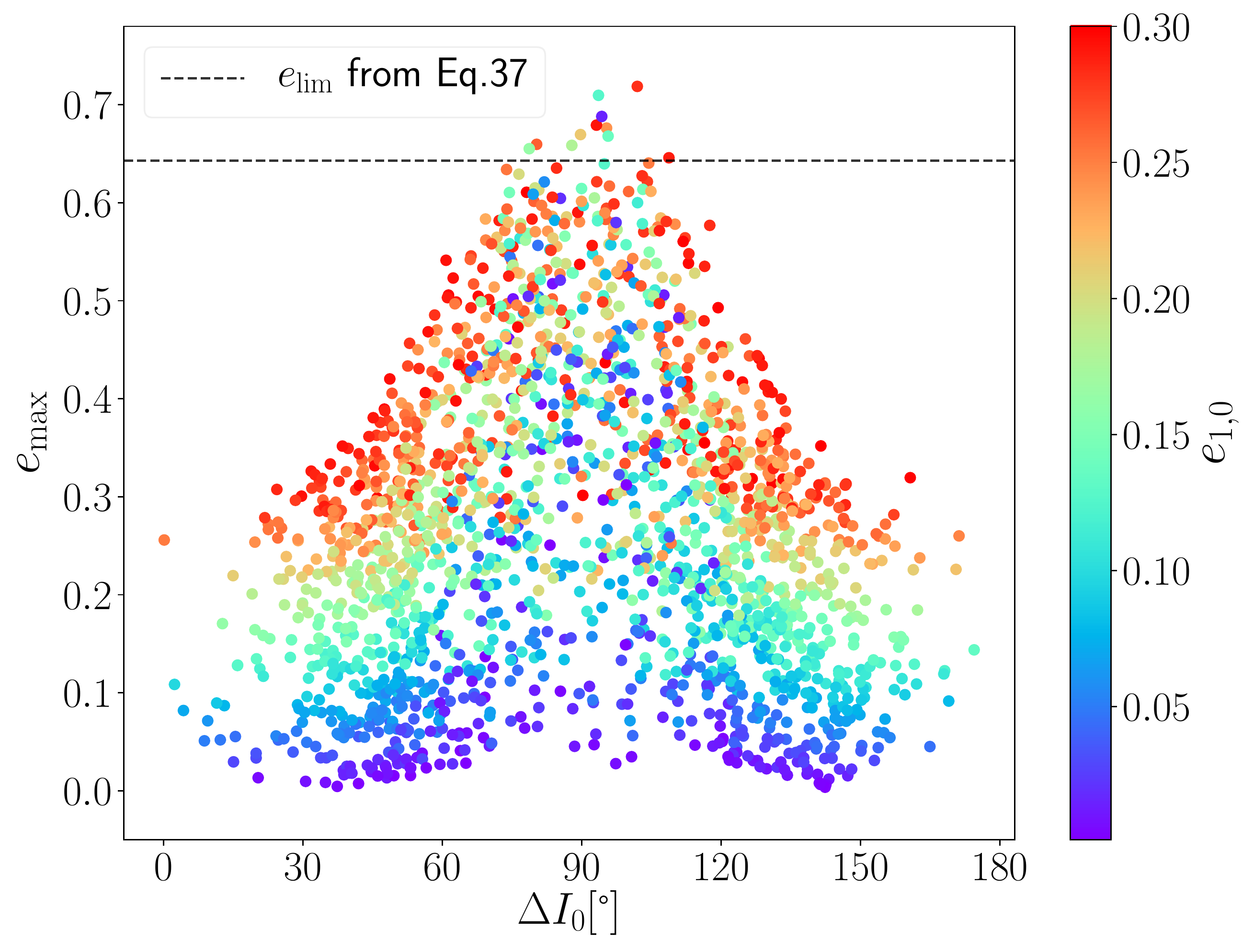}
    \caption{The maximum eccentricity reached in each simulation, plotted as a function of the initial $\Delta I$. The colour represents the initial $e_1$ of the simulation, which is randomly drawn between $0-0.3$. The dashed line shows the limiting eccentricity predicted from Eq.~\ref{eq:elim} for our system, which is derived analytically in \citet{liu_black_2019}.}
    \label{fig:e_max_sim}
\end{figure}

In Fig.~\ref{fig:e_max_sim}, we plot $e_{\rm{max}}$ attained in each of our 2000 simulations as a function of the initial $\Delta I$ ($\Delta I_0$). The initial $e_1$ ($e_{1,0}$) is indicated by the colour. From Fig.~\ref{fig:e_max_sim}, we see that higher eccentricities are reached when $\Delta I_0$ is larger. This general behaviour is expected from Eq.~\ref{eq:kl_emax}, which gives $e_{\rm{max}}$ in the TPQ approximation with $e_{1,0}=0$ (and no short-range forces). As shown by the colour differences, $e_{\rm{max}}$ also rises with $e_{1,0}$, especially when $\Delta I_0$ is farther away from $90\degr$. Near $90\degr$, the colours become mixed, meaning that even configurations with low $e_{1,0}$ can reach high values of $e_{\rm{max}}$.

Both \citet{fabrycky_shrinking_2007} and \citet{liu_suppression_2015} have estimated the maximum eccentricity attainable when large eccentricity oscillations in the regime $\Delta I=39.2\degr-140.8\degr$ are suppressed by a faster GR precession rate, and found that $e_{\rm{max}}$ is a function of $\Delta I_0$ and $e_{1,0}$, as well as the relative sizes of $\tau_{\rm{quad}}$ and $\tau_{\rm{GR}}$. We parametrize these time scales in a similar manner as \citet{liu_suppression_2015} to define $\epsilon_{\rm{GR}}$
\begin{equation}\label{eq:epi_gr}
    \epsilon_{\rm{GR}} = \frac{\tau_{\rm{quad}}}{\tau_{\rm{GR}}} (1-e_1^2). 
\end{equation}
$\tau_{\rm{quad}}$ and $\tau_{\rm{GR}}$ are given in Eq.~\ref{eq:t_kl} and Eq.~\ref{eq:t_gr}, respectively. The term $(1-e_1^2)$ is cancelled out by the same term in $\tau_{\rm{GR}}$, so $\epsilon_{\rm{GR}}$ does not depend on $e_1$. For $\pi$ Men c, $\epsilon_{\rm{GR}} \approx 2.1$. Using $\epsilon_{\rm{GR}}$, we can estimate the maximum possible $e_{\rm{max}}$, which is attained when $\Delta I_0 = 90\degr$. This is termed the limiting eccentricity ($e_{\rm{lim}}$) in \citet{liu_suppression_2015}. Using eq.3 in \citet{liu_black_2019}, and simplifying with $L_1 \ll L_2$, we get
\begin{equation}\label{eq:elim}
    0 = \frac{3}{8} (3 e_{\rm{lim}}^2 + 2 e_{1,0}^2) + \epsilon_{\rm{GR}} \left(\frac{1}{\sqrt{1-e_{1,0}^2}}-\frac{1}{\sqrt{1-e_{\rm{lim}}^2}} \right),
\end{equation}
where $e_{1,0}$ is the initial eccentricity of the inner orbit. Solving this with $e_{1,0} = 0.3$, our maximum starting $e_1$, we get $e_{\rm{lim}} \approx 0.64$. We overplot $e_{\rm{lim}}$ in Fig.~\ref{fig:e_max_sim}, and find that the majority of simulations are below $e_{\rm{lim}}$, in agreement with the analytic calculations. The few outliers are expected due to the limited precision of secular averaging compared to $N$-body simulations. In summary, the excitation of the inner eccentricity is limited to a value of $\approx 0.64$, which would occur at nearly $90\degr$ mutual inclinations between the inner and outer orbit, a configuration that is perhaps unlikely in nature. At most other values of $\Delta I$, $e_{\rm{max}}$ attains moderate values (e.g. $e_{\rm{max}}< 0.5$ for about 92 per cent of simulations).

We note that Eq.~\ref{eq:elim} assumes no damping in $e$, and hence tends to over-estimate $e_{\rm{lim}}$. As mentioned in \S\ref{sec:tidal_diss}, the tidal dissipation efficiency is highly uncertain for $\pi$ Men c. If $\tau_e$ is less than the system age, one should include the effect of tides when interpreting the eccentricity evolution. For example, if tides have damped away the free eccentricity, the actual $e_{\rm{max}}$ would likely tend to a value close to one of the purple dots in Fig.~\ref{fig:e_max_sim} (i.e. simulations where $e_{1,0}\approx0$). However, it is also possible that $\tau_e$ is longer than the system age and eccentricity damping is negligible. To recognize this uncertainty, we will draw attention to instances where $e_1$ is used in our calculations, specifically in Eq.~\ref{eq:Omega_precess} and Eq.~\ref{eq:transit}.

\begin{figure}
    \centering
    \includegraphics[width=0.8\linewidth]{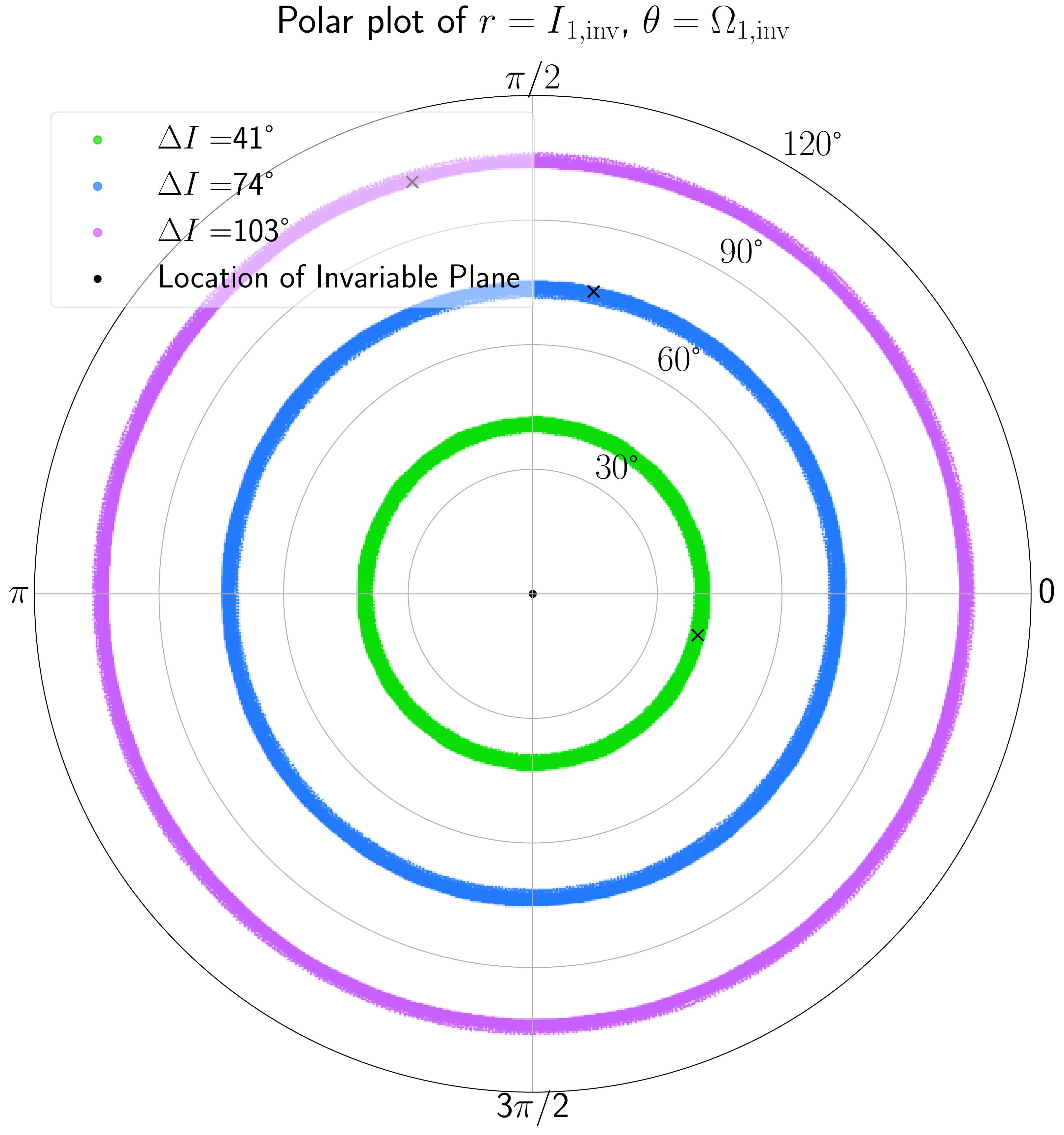}
    \caption{Polar plot of $r=I_{\rm{1,inv}}$, $\theta=\Omega_{\rm{1,inv}}$. The trajectories start at the crosses, and precess around the origin, with a radial size set by $\Delta I \approx I_{\rm{1,inv}}$. The two $\Delta I < 90\degr$ trajectories precess clockwise, while the $\Delta I > 90\degr$ trajectory precesses counterclockwise. Simulations with three different $\Delta I$ are plotted. For each $\Delta I$, we run a series of 36 simulations with different initial values of $e_1$ (four evenly-spaced values from $0-0.3$) and $\omega_1$ (nine evenly-spaced values from $0-2\pi$). The finite width of the lines show that $I_{\rm{1,inv}}$ can evolve by a few degrees over the course of a simulation. Simulations used in this plot are not part of the final set of 2000 simulations, but ran for illustrative purposes.}
    \label{fig:polar_inv}
\end{figure}

\subsection{Nodal precession driven by the outer planet}\label{sec:Omega_precess}
In this subsection, we examine nodal precession of the inner planet, $\pi$ Men c, due to interactions with the outer planet, $\pi$ Men b. In reality, the quadrupole moment of the rotating and oblate star also induces precession. Using the results of \citet{lai_how_2018}, however, we find that the precession rate due to stellar oblateness is negligibly small for $\pi$ Men c compared to that induced by the outer planet (see \S\ref{sec:geom_overview} for details). Therefore, in this subsection, we only consider precession of the inner orbit around the total angular momentum vector of the two planets, which is very similar to the outer planet's angular momentum vector.

We can visualize nodal precession by examining the $\Omega_{\rm{inv}}$ and $I_{\rm{inv}}$ phase space for the inner planet. As shown in Fig.~\ref{fig:polar_inv}, the planet traces out cyclical trajectories in this phase space as it precesses. Trajectories with different colours correspond to simulations with a different $\Delta I$, which sets the size of the trajectory. We plot the results from a series of simulations with varying $e_1$ and $\omega_1$ for each $\Delta I$, which explains the finite width of the lines, and shows that $I_{\rm{1,inv}}$ can oscillate by a few degrees (see also bottom left panel of Fig.~\ref{fig:evol_all}).

The time scale of nodal precession can be estimated from Eq.~\ref{eq:Omega_dot}, which gives the time derivative of $\Omega$. Since large eccentricity oscillations are suppressed in this system, we follow \citet{yee_hat-p-11_2018} and drop the term with $\cos{2\omega_1}$ from the Hamiltonian (see Eq.~\ref{eq:Hamiltonian} and Eq.~\ref{eq:f_quad_delaunay}). Taking the time derivative of remaining Hamiltonian, we get
\begin{equation}\label{eq:Omega_precess}
\begin{split}
    \dot{\Omega}_1 = -\frac{\partial \curH}{\partial H_1} \\
    &= -\frac{1}{8} \sqrt{\frac{G}{M_\star}} \left(\frac{a_1^{3/2}}{a_2^{3}}\right) \frac{m_2}{(1-e_2^2)^{3/2}} \left(\frac{15H_1}{G_1^2}-9H_1\right), \\
\end{split}
\end{equation}
where $H_1$ and $G_1$ are given in Eq.~\ref{eq:omega_dot} and Eq.~\ref{eq:Omega_dot}. $H_1$ is a function of $I_1$, which is approximately equal to $\Delta I$. Both $H_1$ and $G_1$ are also functions of $e_1$. The sign indicates the direction of precession, and the associated time scale is $P_\Omega = |2\pi/\dot{\Omega}_1|$. Our calculation is intended to serve as a rough estimate. For example, if tidal dissipation reduces the free eccentricity then $P_\Omega$ will be slightly shorter than that calculated here by a factor $\sim\sqrt{1-e_1^2}$. More exact estimates can be performed by considering tides and higher order terms in the perturbing function \citep[e.g.][]{bailey_nodal_2020}.

\begin{figure}
    \centering
    \includegraphics[width=0.9\linewidth]{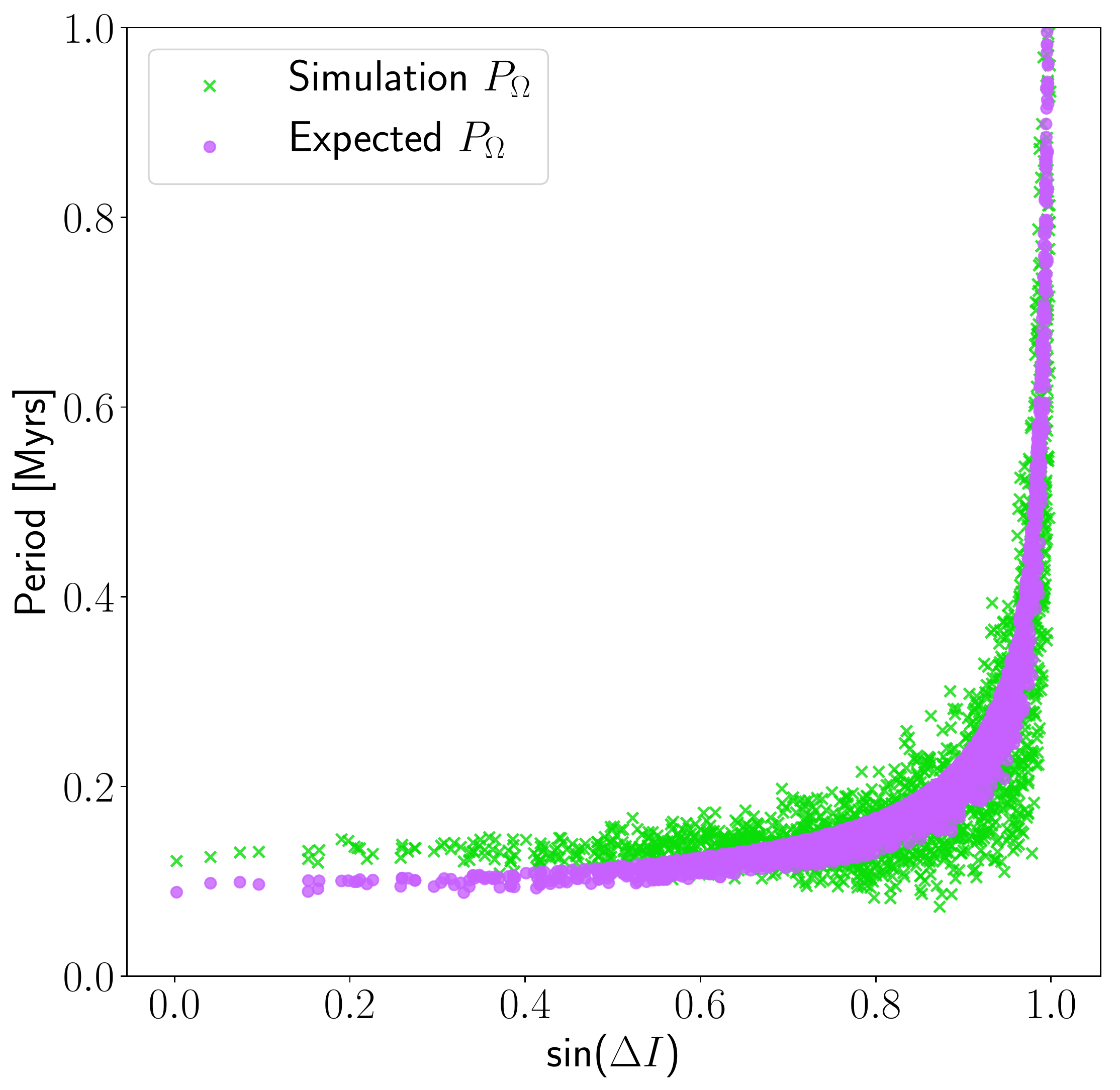}
    \caption{Comparison between the expected $P_{\Omega}$ (purple dots) calculated from Eq.~\ref{eq:Omega_precess} and the actual $P_{\Omega}$ (green crosses) from simulations. We truncate the $y$-axis to 1 Myr for demonstration purposes, and note that the agreement gets worse as the expected $P_{\Omega} > 1$ Myr, corresponding to $ 85\degr \lesssim \Delta I \lesssim 95\degr$.
    \label{fig:P_Omega}}
\end{figure}

In Fig.~\ref{fig:P_Omega}, we plot the expected $P_\Omega$ from Eq.~\ref{eq:Omega_precess} as a function of $\sin{\Delta I}$, as well as the actual $P_\Omega$ derived from the simulations. Since Eq.~\ref{eq:Omega_precess} is a function of $e_1$, we use the median value of $e_1$ from each simulation to compute the expected time scale. To estimate the actual $P_\Omega$ from the simulations, we use a Lomb-Scargle periodogram to search for periodic signals in $\Omega$ within the range $0.1 - 10$ Myr. As shown in Fig.~\ref{fig:P_Omega}, we find good agreement between the expected and actual $P_\Omega$, especially at lower $\sin{\Delta I}$, or $\Delta I$ away from $90\degr$. As $\Delta I$ approaches $90\degr$, the precession time scale goes to infinity since Eq.~\ref{eq:Omega_precess} goes to zero. 

We note that nodal precession has a longer time scale than the eccentricity excitation time scale shown in Fig.~\ref{fig:evol_all}, which is dominated by the GR precession time scale. This is also apparent in the different periods of $\omega_1$ and $\Omega_1$ shown in Fig.~\ref{fig:evol_all}. In the next section, we examine some interesting observational consequences of nodal precession.

\section{Effects of nodal precession in large $\Delta I$ systems}\label{sec:nod_prec_effects}
In this section, we explore two consequences of nodal precession in systems like $\pi$ Men and HAT-P-11, which have a large planet-planet $\Delta I$. First, as the inner planet undergoes nodal precession around the invariable plane, which is misaligned to its orbit by $\approx \Delta I$, the stellar obliquity ($\psi$, the angle between the stellar spin axis and orbital axis of inner planet) could evolve significantly as long as the stellar spin axis is not aligned with the outer planet's orbital axis, and the star and inner planet are not strongly coupled. Second, the inner planet comes in and out of a transiting configuration as it precesses. After discussing the relative precession rates and outlining the geometry in \S\ref{sec:geom_overview}, we discuss these two effects in \S\ref{sec:stellar_spin} and \S\ref{sec:p_dyn}, respectively.

\subsection{Relative precession rates and geometry}\label{sec:geom_overview}
To understand these effects, consider three relevant directions: the orientations of the two orbital planes, $\Lone$ and $\Ltwo$, and the orientation of the stellar spin axis, $\Lstar$. Each of the three vectors precesses due to the quadrupole moment of the other two bodies. In \S\ref{sec:Omega_precess}, we considered the precession of $\Lone$ around $\Ltwo$ as driven by the quadrupole moment of $m_2$. Adopting the framework of \citet{lai_how_2018}, we now consider if other precession cycles may be important in the system. As a simplification, we assume that $\Ltwo$ is fixed and traces the invariable plane in both $\pi$ Men and HAT-P-11. Besides the precession rate of $\Lone$ around $\Ltwo$, which can be termed $\omega_{12}$, there are three other important precession rates in the system, $\omega_{1\star}$, which is the precession rate of the $\Lone$ around $\Lstar$, $\omega_{\star1}$, the precession rate of $\Lstar$ around $\Lone$, and $\omega_{\star2}$, the precession rate of $\Lstar$ around $\Ltwo$. 

These rates can be compared to determine the dominant behaviour of the system. We are specifically interested in the evolution of $\psi$ and $\Lone$ over time. \citet{lai_how_2018} showed that the time evolution of $\psi$ has three different regimes, as determined by the parameter \citep[eq. 8 of][]{lai_how_2018}
\begin{equation}
    \epsilon_{\star1} \equiv \frac{\omega_{12} - \omega_{\star2}}{\omega_{\star1}+\omega_{1\star}},
\end{equation}
which measures the coupling strength between the star and the inner planet relative to the forcing of the outer planet. If $\epsilon_{\star1} \ll 1$, the mutual precession between $\Lstar$ and $\Lone$ is faster than $\omega_{12}$, meaning that the star and inner planet are strongly coupled so that $\Lone$ and $\Lstar$ follow each other. In this case, the obliquity cannot evolve significantly \citep{lai_how_2018}. As we discuss below, $\psi$ does evolve significantly for the regime where we find $\pi$ Men and HAT-P-11 to be in.

Lastly, we note that tidal dissipation could affect $\psi$. Applying the results of \citet{barker_tidal_2009} to $\pi$ Men c, however, we find that the time scale to align the stellar spin with the inner orbit is comparable to $\tau_a$, which we argue in \S\ref{sec:tidal_diss} must be much longer than the ages of $\pi$ Men and HAT-P-11 given that their inner planets are still at relatively large orbital distances. We therefore assume that tidal dissipation is negligible for the evolution of $\psi$ in this paper.

\subsubsection{Weak planet-star coupling: $\pi$ Men and HAT-P-11}\label{sec:weakcoup_pimen}
If $\epsilon_{\star1} \gg 1$, then the inner planet and the star are weakly coupled. This is true when $\omega_{\star1}$ and $\omega_{1\star}$ are much smaller than $\omega_{12}$. In this regime, $\Lone$ precesses around $\Ltwo$ at a rapid rate, while $\Lstar$ changes slowly. We find that this is the case for both $\pi$ Men and HAT-P-11.\footnote{To estimate $\omega_{1\star}$ and $\omega_{\star1}$, we assume a stellar rotation period of $P_\star=18.3$ d for $\pi$ Men \citep{zurlo_imaging_2018} and $P_\star=29.2$ d for HAT-P-11 \citep{bakos_hat-p-11b_2010}, stellar mass and radius values from \citet{huang_tess_2018} and \citet{yee_hat-p-11_2018}, respectively, and typical constants for the quadrupole moment of a solar-type star from \citet{lai_how_2018}.} Specifically, for $\pi$ Men we find $\epsilon_{\star1} \approx 36$, and $\omega_{12} \gg \omega_{1\star} \gg \omega_{\star1} \gg \omega_{\star2}$, while for HAT-P-11 we find $\epsilon_{\star1} \approx 7$, and $\omega_{12} \gg \omega_{\star1} > \omega_{1\star} \gg \omega_{\star2}$.

\begin{figure}
    \centering
    \includegraphics[width=0.85\linewidth]{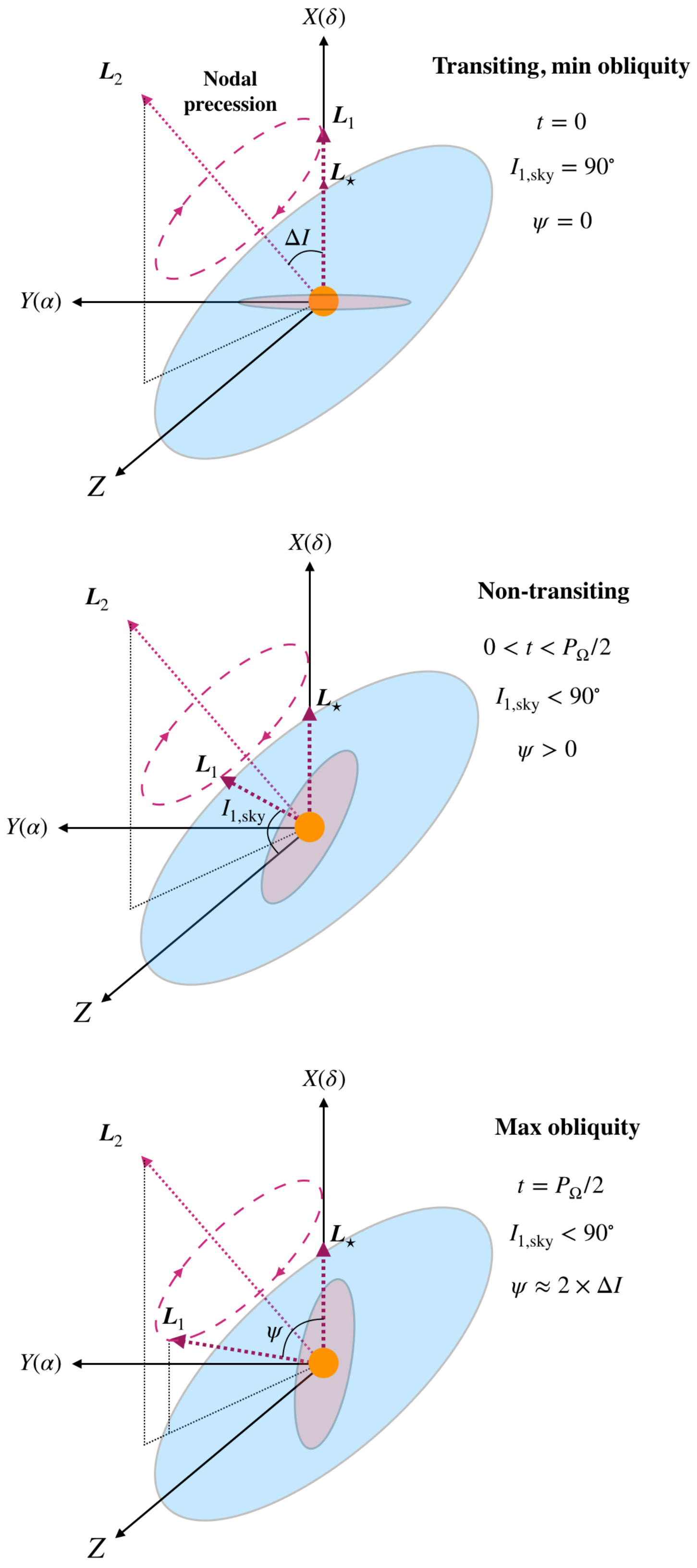}
    \caption{Illustration of nodal precession and the resultant evolution of $I_{\rm{1,sky}}$ and $\psi$ for the case of weak coupling between the inner planet and the star. Here, $X$ and $Y$ define the sky plane, and are along the $\delta$ and $\alpha$ directions respectively. $Z$ points toward the observer. The stellar spin axis $\boldsymbol{L}_\star$ points in $X$, and $\boldsymbol{L}_2$ points out of the sky plane. Both $\boldsymbol{L}_\star$ and $\boldsymbol{L}_2$ can be considered fixed in the weak coupling case. At $t=0$ (top panel), $\boldsymbol{L}_1$ points in the $X$ direction and $I_{\rm{1,sky}}=90\degr$, $\psi=0$. $\boldsymbol{L}_1$ then precesses around $\boldsymbol{L}_2$, causing $\boldsymbol{L}_1$ to come out of the sky plane (middle panel). The obliquity has increased and the inner planet no longer transits. At $t=P_\Omega/2$, $\psi$ reaches its maximum value of $\approx 2 \times \Delta I$, and the inner planet is still in a non-transiting configuration.}
    \label{fig:illustrate_nodal}
\end{figure}

We illustrate the evolution of $\Lone$ for our two systems in Fig.~\ref{fig:illustrate_nodal}. For simplicity, we assume that $\Lone$ and $\Lstar$ are initially both in the sky plane ($X-Y$ plane), along the $X$ direction. In other words, $\psi = 0$ and $I_{\rm{1,sky}}=90\degr$ initially (top panel). $\Ltwo$ points out of the sky plane, in a different direction than $\Lstar$. Since the fastest frequency is $\omega_{12}$, the dominant behaviour is that $\Lone$ precesses in a circle (dashed circle in Fig.~\ref{fig:illustrate_nodal}) around $\Ltwo$. $\Lstar$ also precesses around $\Lone$ and $\Ltwo$, but at much slower rates, so we ignore this precession in Fig.~\ref{fig:illustrate_nodal} and assume $\Lstar$ is fixed. As $\Lone$ precesses around $\Ltwo$, $\Lone$ comes out of the sky plane (middle panel). Initially, the inner planet is seen transiting. However, as $\Lone$ comes out of the sky plane, $I_{\rm{1,sky}}$ changes and the inner planet may not transit. The larger the mutual inclination, the more $I_{\rm{1,sky}}$ can deviate from $90\degr$, so the more often the inner planet will not transit. At the same time, $\psi$ is changing since the relative orientation between $\Lone$ and $\Lstar$ is varying with time. If at $t=0$, $\psi=0$, then $\psi$ can reach a maximum value of $\approx 2 \times \Delta I$ after half the nodal precession period (bottom panel), a result that was also derived analytically by \citet{lai_how_2018}. For HAT-P-11, this means that $\Delta I \approx 50\degr$ between the two planets would be sufficient to produce $\psi\approx100\degr$.

We emphasize that the change in obliquity here is dependent on the misalignment between $\Ltwo$ and $\Lstar$. In the case where $\Ltwo$ and $\Lstar$ point in the same direction, the obliquity would stay roughly constant and be equal to $\Delta I$. It will not be completely constant because the dashed circle of precession has a finite width of a few degrees, which corresponds to the small variation in $I_{\rm{1,inv}}$ over the course of a simulation (see Fig.~\ref{fig:polar_inv}).

\subsubsection{Summary of possible configurations}
The discussion surrounding Fig.~\ref{fig:illustrate_nodal} involves two possible configurations of the system. In general, we can envision three different configurations of $\Lone$, $\Lstar$, and $\Ltwo$.

\begin{itemize}
    \item[] I. ($\Lone \approx \Lstar \ne \Ltwo$): The orbit of the inner planet and the stellar spin axis are aligned, but the outer planet's orbit is misaligned with both. This is the initial state of the system in Fig.~\ref{fig:illustrate_nodal}.
    \item[] II. ($\Lone \ne \Lstar \approx \Ltwo$): The orbit of the outer planet is aligned with the stellar spin direction, while the inner planet is misaligned with both. In this case, $\psi \approx \Delta I$.
    \item[]  III. ($\Lone \ne \Lstar \ne \Ltwo$): None of the three angular momentum vectors are aligned in the system. This is the long-term state of the system in Fig.~\ref{fig:illustrate_nodal}.
\end{itemize}

What we find is that due to nodal precession, an alignment between $\Lone$ and $\Lstar$ is very fortuitous in systems like $\pi$ Men and HAT-P-11. Therefore, case I is rare and large $\Delta I$ systems will usually be observed in case II or case III. We will argue in \S\ref{sec:diss_formation} that due to dynamical considerations, case III is more likely for $\pi$ Men and HAT-P-11 than case II, so the stellar obliquity will likely evolve significantly with time in these systems.

\subsection{Measurements of the stellar spin orientation} \label{sec:stellar_spin}
Given the above findings, observational constraints on the stellar spin orientation in $\pi$ Men and HAT-P-11 would be helpful. We summarize constraints from previous observations below.

\subsubsection{$\pi$ Men}
For $\pi$ Men, we can derive a rough constraint on $\Lstar$ using previous measurements of the rotational period ($P_\star$) and projected rotational velocity ($v \sin{I_\star}$). Our data for $P_\star$ comes from \citet{zurlo_imaging_2018}, who used archival time series photometry from the All Sky Automated Survey to detect a periodic signal of $18.3\pm1$ d ($99$ per cent confidence). We assume that the 18.3 d signal coincides with the stellar rotation period. To translate $P_\star$ into the rotational velocity $v$, we use the radius measurement of $R_\star = 1.10\pm0.023~\Rsun$ from \citet{huang_tess_2018}. There are multiple $v \sin{I_\star}$ measurements for $\pi$ Men, and most are consistent. We adopt $v \sin{I_\star}=2.96\pm0.28~{\mathrm{km}\,\mathrm{s}^{-1}}$ from \citet{delgado_mena_li_2015}. 

With these measurements, we perform a MCMC sampling with uniform priors in $v$ and $\cos{I_\star}$, following the statistically correct procedure described in \citet{masuda_inference_2020-1}. We find that $I_\star = 74.6\degr^{+10.5\degr}_{-12.1\degr}$ ($1\sigma$). Given that $I_{\rm{1,sky}} = 87.46\pm0.08\degr$, this implies that $\pi$ Men c could be consistent with being well-aligned with its star (i.e. case I). An alignment would not be in conflict with the large $\Delta I$ between the planets, as Fig.~\ref{fig:illustrate_nodal} shows that $\psi$ varies cyclically, and the exact value depends on where the planet happens to be in the cycle, as well as the initial value of $\psi$. However, an alignment between $\Lone$ and $\Lstar$ would require us to be observing the system at a special time, and the current data alone is also consistent with the star being aligned with the orbit of the outer planet (i.e. $\Ltwo \approx \Lstar$), which has $I_{\rm{2,sky}}=41-65\degr$ (i.e. case II). Finally, since these sky-projected inclinations do not probe the exact orientations, it is also possible that $\Lone \ne \Lstar \ne \Ltwo$, giving case III.

Therefore, for $\pi$ Men, current observations cannot rule out any of the three cases outlined in the previous subsection. A direct measurement of $\lambda>0$ from the Rossiter-Mclaughlin effect could eliminate case I. Observing $\lambda \approx \psi \approx \Delta I$ would provide indirect evidence for case II. Finally, asteroseismology would provide a more precise estimate of $I_\star$.

\subsubsection{HAT-P-11}
As mentioned in \S\ref{sec:hatp_I}, \citet{yee_hat-p-11_2018} invoke nodal precession as a possible cause of the large sky-projected obliquity of HAT-P-11 b ($\lambda \approx 100\degr$, a proxy for the true obliquity $\psi$). In \S\ref{sec:hatp_I}, we found that $54\degr < \Delta I < 126\degr$ ($1 \sigma$) for HAT-P-11, which is sufficient to produce $\psi \approx 100\degr$.

The high obliquity observed in HAT-P-11 means that $\Lone \ne \Lstar$. Our $\Delta I$ measurement gives $\Lone \ne \Ltwo$. The only remaining uncertainty in HAT-P-11 is whether $\Ltwo$ and $\Lstar$ are aligned. We cannot repeat the method above to constrain $I_\star$ due to the slow and poorly constrained rotation rate of the star ($v \sin{I_\star}=1.5\pm1.5~{\mathrm{km}\,\mathrm{s}^{-1}}$ from \citealt{bakos_hat-p-11b_2010}). Instead, we use previous star-spot crossing observations for HAT-P-11 \citep{sanchis-ojeda_starspots_2011-1}, which find a degeneracy between nearly pole-on ($I_\star \sim 170\degr$) and nearly edge-on ($I_\star \sim 80\degr$) configurations of the stellar spin axis. If $I_\star \sim 80\degr$ and $I_\star \sim 170\degr$ are the only two possibilities, this suggests that the star would be misaligned with the outer planet as well, which has $114\degr < I_{\rm{2,sky}} < 148\degr$ or $34\degr < I_{\rm{2,sky}} < 70\degr$ (both $1\sigma$ intervals). Therefore, it appears that none of the three vectors are aligned in HAT-P-11, giving case III ($\Lone \ne \Lstar \ne \Ltwo$).

\subsection{The dynamical transit probability of $\pi$ Men c}\label{sec:p_dyn}
As nodal precession occurs, $\pi$ Men c shifts in and out of a transiting configuration (see Fig.~\ref{fig:illustrate_nodal}). The time-varying nature of transits due to nodal precession has been noted by past studies \citep[e.g.][]{martin_planets_2014,martin_transit_2017, boley_transit_2020}, especially for circumbinary planets for which the precession time scales can be as short as tens of years (e.g. {\it Kepler} 413 b, \citealt{kostov_kepler-413b_2014}; {\it Kepler} 1661 b, \citealt{socia_kepler-1661_2020}). Even a small misalignment between the planetary orbit and the binary orbit will cause circumbinary planets to shift in and out of a transiting state, similar to what we find for $\pi$ Men c where in this case the binary is a planetary-mass companion. Recently, one planet (K2-146 c) was also found to come into transit from a non-transiting configuration due to an unknown perturber \citep{hamann_k2-146_2019}. 

From our simulations, we can determine quantitatively how often $\pi$ Men c transits in the long term. We term this the dynamical transit probability, $P_{\rm{dyn}}$. This should not be confused with the more common geometric transit probability, $P_{\rm{geom}}$, which describes how likely it is see a planet transiting if the distribution of possible orbits is isotropic. Taking into account the enhanced probability of seeing eccentric planets, $P_{\rm{geom}}$ is given by \citep{borucki_photometric_1984,barnes_effects_2007}
\begin{equation}
    P_{\rm{geom}} = \frac{\Rstar}{a_p (1-e^2)}.
\end{equation}
Using $a_p / R_\star \approx 13.4$ \citep{huang_tess_2018}, we get $P_{\rm{geom}} \approx 7.5$ per cent for a circular orbit of $\pi$ Men c, which corresponds to a lower limit of $P_{\rm{geom}}$. On the other hand, $P_{\rm{dyn}}$ measures how often a planet is seen transiting in the long term, taking into account secular interactions that render its possible orbits not completely isotropic. While $P_{\rm{geom}}$ is statistically useful in population-level studies (e.g. deriving planet occurrence rates), $P_{\rm{dyn}}$ should only be considered on a case-by-case basis. 

\begin{figure}
    \centering
    \includegraphics[width=0.8\linewidth]{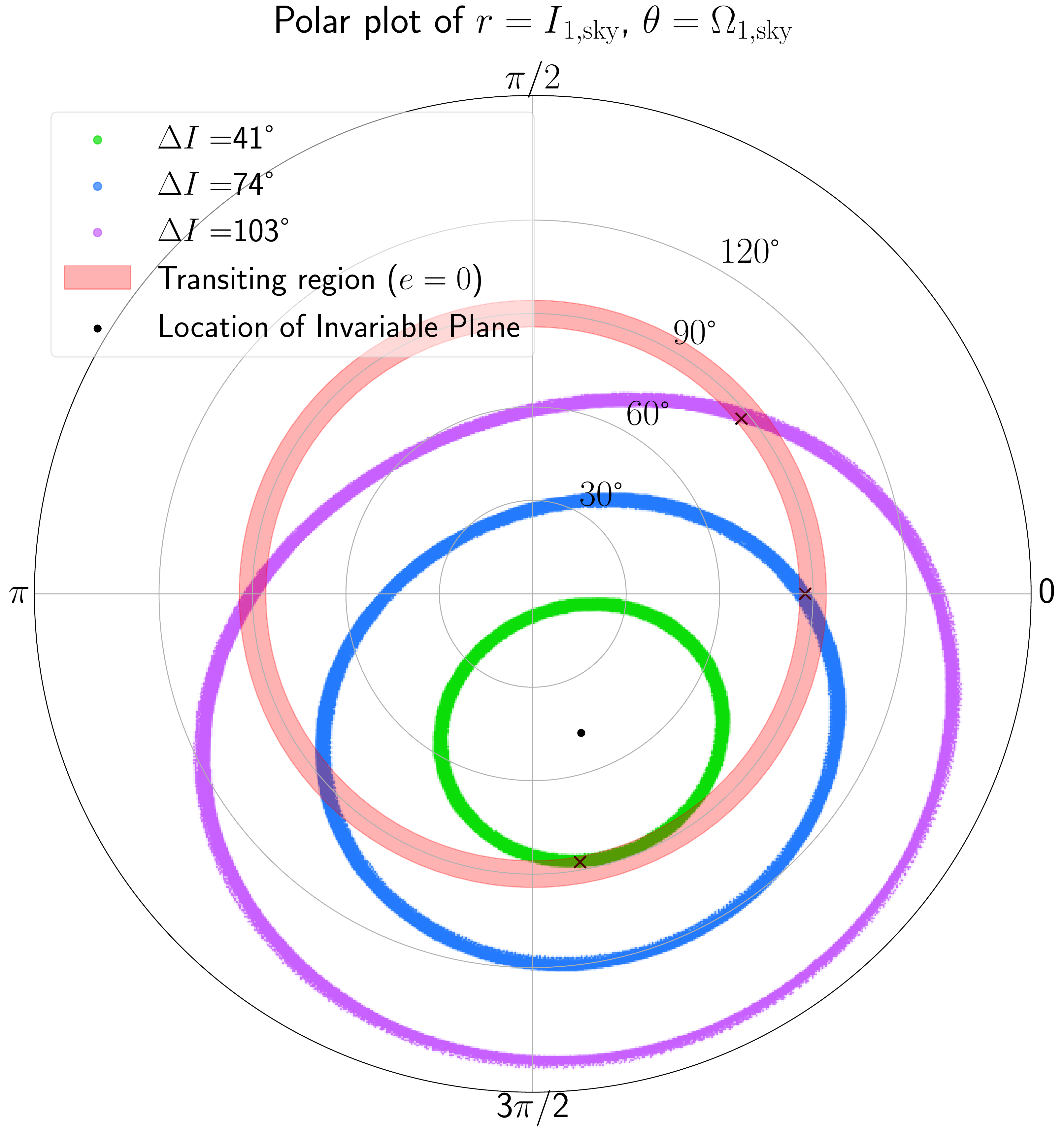}
    \caption[]{Polar plot of $r=I_{\rm{1,sky}}$, $\theta=\Omega_{\rm{1,sky}}$. The same simulations as in Fig.~\ref{fig:polar_inv} are now represented in the sky plane. The transiting region is shown as a band of width $I_{\rm{crit}} = 4.3\degr$ centred on $90\degr$, where this particular $I_{\rm{crit}}$ corresponds to circular orbits of the inner planet. The location of the invariable plane is plotted as the black dot. For clarity, we make the invariable plane similar in these three cases. The trajectories start at the crosses, and proceed clockwise (if $\Delta I < 90\degr$) or counterclockwise (if $\Delta I > 90\degr$) around the invariable plane.}
    \label{fig:polar_sky}
\end{figure}

To estimate $P_{\rm{dyn}}$ for $\pi$ Men c, we convert our orbital angles back to the sky plane. In the sky plane, we can estimate the size of the transiting region in terms of the maximum angle that $I_{\rm{1,sky}}$ can be away from $90\degr$ (i.e. perfectly edge-on). We call this maximum angle $I_{\rm{crit}}$. The planet transits if $I_{\rm{1,sky}}$ is between $90\degr - I_{\rm{crit}}$ and $90\degr + I_{\rm{crit}}$. For circular orbits, $I_{\rm{crit}} = \arctan(R_\star/a_p)$, or $\approx$ $4.3\degr$ for $\pi$ Men c. However, secular perturbations give $\pi$ Men c a non-zero eccentricity, which enhances its transit probability at a given time. We take this enhancement into account by widening the effective transiting band by a factor of $1/(1-e^2)$, where $e$ is taken to be the eccentricity at that time \citep{barnes_effects_2007}. In summary, we determine the planet to be transiting if
\begin{equation}
    90-\frac{\arctan(R_\star/a_p)}{(1-e^2)} \leq I_{\rm{1,sky}} \leq 90+\frac{\arctan(R_\star/a_p)}{(1-e^2)}.
    \label{eq:transit}
\end{equation}
In Fig.~\ref{fig:polar_sky}, we show polar plots of $r=I_{\rm{1,sky}}$, $\theta=\Omega_{\rm{1,sky}}$. This is similar to Fig.~\ref{fig:polar_inv}, which showed the two angles in the invariable plane. Overplotted in red in Fig.~\ref{fig:polar_sky} is the transiting band, assuming circular orbits (the actual size of the region changes as a function of $e_1$). The inner planet traces out cyclical trajectories in this space, so one can easily visualize when the planet is transiting.

As shown in Fig.~\ref{fig:polar_sky}, the inner planet is not always in transit, and the exact $P_{\rm{dyn}}$ depends on the initial conditions of the simulation. We estimate $P_{\rm{dyn}}$ for each simulation by finding the fraction of timesteps that $\pi$ Men c is transiting over one complete cycle in $\Omega$. For each timestep, we use the prescription in Eq.~\ref{eq:transit} to determine if the planet is transiting.\footnote{As we do not include tidal dissipation in our simulations, our values of $e$ would be larger on average than what is expected if tidal dissipation has damped away the free eccentricity. Therefore, we may slightly over-estimate $P_{\rm{dyn}}$ by less than 1 per cent.} We use only one nodal precession period to avoid biasing $P_{\rm{dyn}}$ by unevenly sampling the long-term cycle. The time scale of nodal precession is relatively long, especially for $\Delta I$ close to $90\degr$, so our simulations are run long enough to sample at least one cycle in $\Omega$.

\begin{figure}
    \centering
    \includegraphics[width=0.85\linewidth]{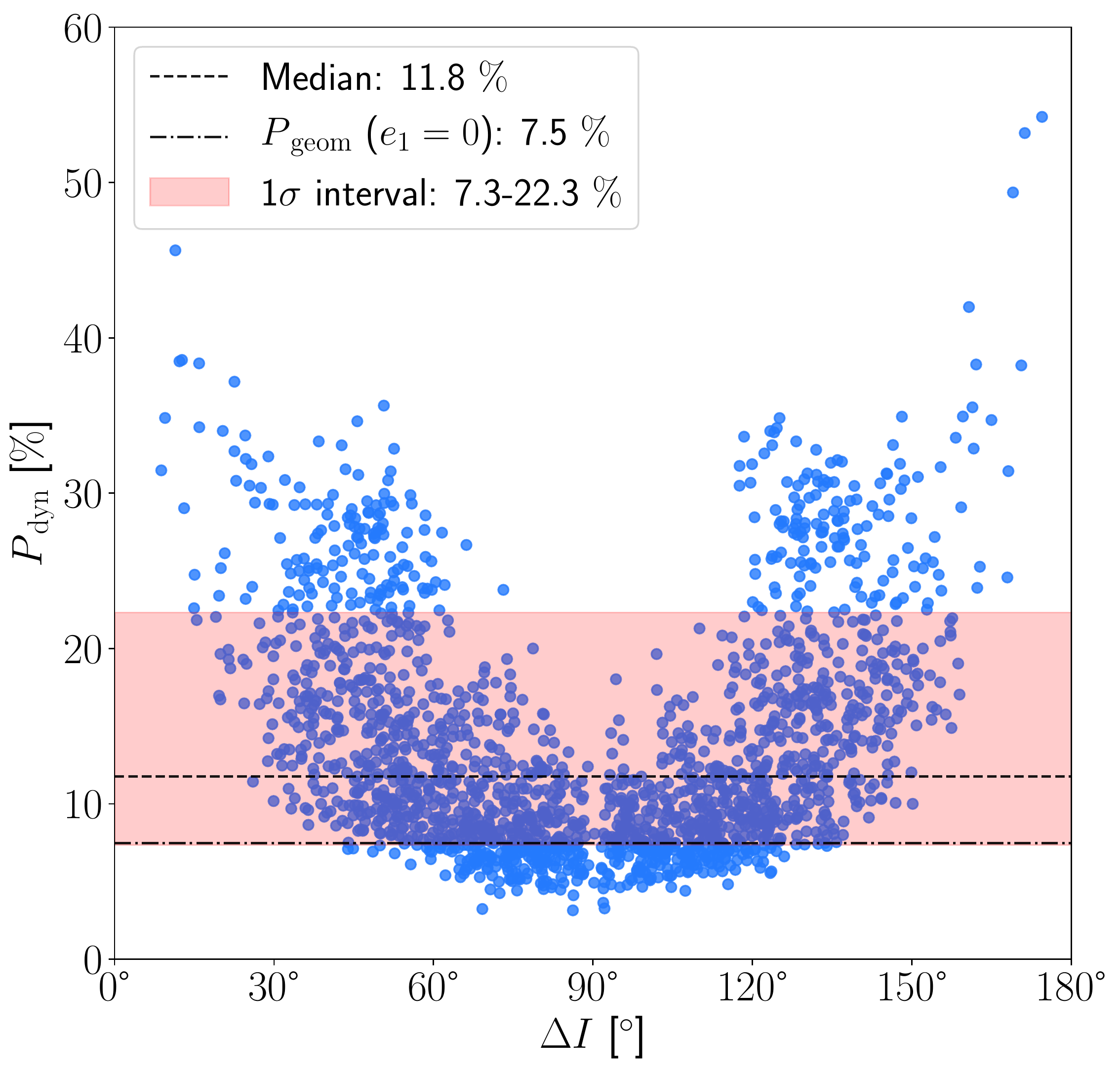}
    \caption{$P_{\rm{dyn}}$ of $\pi$ Men c as a function of the initial $\Delta I$ from our simulations. The $1\sigma$ interval is traced by the light red region, while the dashed line indicates the median. The value of $P_{\rm{geom}}$ for a circular inner orbit is overplotted as the dash-dotted line.}
    \label{fig:Pdyn}
\end{figure}

We plot the estimated $P_{\rm{dyn}}$ from our full set of 2000 simulations in Fig.~\ref{fig:Pdyn}, which shows that $P_{\rm{dyn}}$ is between $7-22$ per cent (1$\sigma$ interval), with a median at $12$ per cent, and decreases as $\Delta I$ gets closer to 90$\degr$. The $P_{\rm{dyn}}$ of $\pi$ Men c is comparable to that of {\it Kepler} 1661 b, which is found to only transit $\sim7$ per cent of the time \citep{socia_kepler-1661_2020}. We also overplot the value of $P_{\rm{geom}}$ for $e_1=0$ in Fig.~\ref{fig:Pdyn} for comparison. Given the distribution of mutual inclinations we derive, $\pi$ Men is in fact not transiting a majority of the time. Our derived values for $P_{\rm{dyn}}$ are generally higher than the circular $P_{\rm{geom}}$ (7.5 per cent), except for $\approx 15$ per cent of simulations, which have trajectories in the $I_{\rm{1,sky}}-\Omega_{\rm{1,sky}}$ phase space that overlap very little with the transiting region.

Even if there are other non-transiting, super Earth sized planets near $\pi$ Men c, we expect our results on $P_{\rm{dyn}}$ to hold. Depending on the relative coupling between the planets, either the inner planets are strongly coupled and precess in unison around the outer planet's orbit, or they are weakly coupled and precess out of sync (see \citealt{lai_hiding_2017-2} and a different formulation in \citealt{read_transit_2017-1}). Regardless of the coupling strength between the planets, however, $P_{\rm{dyn}}$ will remain low for a given planet. We further discuss the implications of a low dynamical transit probability in \S\ref{sec:Pdyn_discuss}.

\section{Discussion} \label{sec:discuss}
\subsection{Implications of a low dynamical transit probability}\label{sec:Pdyn_discuss}
In \S\ref{sec:p_dyn}, we found that given our inferred $\Delta I$ distribution between $\pi$ Men c and $\pi$ Men b, $P_{\rm{dyn}}$ is low for $\pi$ Men c, on the order of $10-20$ per cent. HAT-P-11 b would also have a similarly low $P_{\rm{dyn}}$ given our measured $\Delta I$ between HAT-P-11 b and c. From our simulations, we estimate that $\pi$ Men will stay transiting for at least two thousand years. The change in its sky-projected inclination is too small to be observable ($\sim 0.001\degr / $  yr), unlike more favourable systems (e.g. Trappist-1 or some circumbinary planets) where it is estimated that planets may de-transit on decade time scales \citep[e.g.][]{boley_transit_2020, socia_kepler-1661_2020}.

The direct implication of the low $P_{\rm{dyn}}$ we infer is that for every transiting $\pi$ Men c, there are about 5-10 other inner planets in misaligned systems like $\pi$ Men that do not transit at the moment. This result does not affect population statistics, which are determined from $P_{\rm{geom}}$, since the distribution of exoplanet orbits as a whole may still be treated as isotropic. Our inferred $P_{\rm{dyn}}$ should be considered in the context of the $\pi$ Men system alone, and not other systems if their $\Delta I$ is unknown.

In the following, we consider the possibility that there are other inner planets in $\pi$ Men that do not currently transit. If this is the case, it might be more likely to see {\it at least} one planet transiting. We denote this probability $P_{\rm{at~least~1}}$. The value of $P_{\rm{at~least~1}}$ depends on the relative coupling between the inner planets. From \citet{lai_hiding_2017-2}, the coupling between two inner planets depends on the relative masses and semi-major axes of the perturber ($\pi$ Men b in this case) and the outermost inner planet (see their eq. 12 \& 38). We note that both \citet{lai_hiding_2017-2} and \citet{read_transit_2017-1} assume low $\Delta I$ ($\leq 10\degr$) between their inner and outer planets, and there could be more complex behaviour at larger $\Delta I$. In general, however, we expect that for $P_{\rm{at~least~1}}$ to increase significantly, the inner planets need to be weakly coupled to each other so that their $P_{\rm{dyn}}$ can be treated as independent and summed to get $P_{\rm{at~least~1}}$. Alternatively, if a resonance between the nodal precession frequencies of the inner planets takes place, there could be large $\Delta I$ between the inner planets \citep{lai_hiding_2017-2}, and this may also enhance $P_{\rm{at~least~1}}$. However, if the inner planets are strongly coupled and precess in unison, $\Delta I$ between them would remain low, in which case it is hard to increase $P_{\rm{at~least~1}}$ by more than $P_{\rm{dyn}}$. The inner system of $\pi$ Men could be in each of these categories (weakly coupled, resonant, strongly coupled) due to the allowed parameter space of additional, non-transiting planets, although the fact that there is only one observed transit could disfavour a strongly coupled scenario (i.e. the one in which we would see either multiple transiting planets or none at all). In summary, in order to increase $P_{\rm{at~least~1}}$ by much more than $P_{\rm{dyn}}$ for $\pi$ Men, there needs to be several other inner planets in $\pi$ Men that have just the right masses and locations. 

\subsection{Dynamically linked inner and outer systems}\label{sec:linked_in_out}

In this section, we consider how outer planets dynamically influence the inner planets in hierarchical systems like $\pi$ Men and HAT-P-11. In particular, past studies have noted a link between dynamically `hot' (eccentric and inclined) outer systems of CJs and dynamically `hot' inner systems of SEs by pointing to correlations between the existence of long-period CJs and inner systems with low transit multiplicities \citep{zhu_super_2018, masuda_mutual_2020}. RV surveys find that long-period CJs have a more or less uniform distribution of eccentricities from 0.0 to 0.8 \citep[e.g.][]{butler_catalog_2006}, so a significant portion are dynamically hot. A low transit multiplicity is a proxy for a dynamically excited inner system because it implies either a low intrinsic multiplicity, e.g. because some planets were destabilized and ejected from the system \citep[e.g.][]{huang_dynamically_2017, mustill_effects_2017}, or a high mutual inclination between the inner planets that reduces the probability of observing multiple transits \citep[e.g.][]{lai_hiding_2017-2, hansen_perturbation_2017, becker_effects_2017, read_transit_2017-1, pu_eccentricities_2018}. The link seems to be causal as well, with outer CJs heating up the inner system. For example, dynamical excitation by an outer planet on the inner system has been proposed to partially explain the {\it Kepler} Dichotomy \citep[e.g.][]{hansen_perturbation_2017}, or the apparent excess of single-transiting systems discovered by {\it Kepler}.

On the other hand, in systems without CJs or systems with CJs that are well-aligned with their inner system, the inner multiplicity is high \citep{zhu_super_2018, masuda_mutual_2020}. In fact, all three systems with transiting CJs used in \citet{masuda_mutual_2020} have three or more inner transiting SEs, and all three CJs are consistent with having low eccentricities, although the uncertainties are large. Our solar system is also a cold and flat system, where the four terrestrial planets and four gas giants have mean inclinations of $\approx 2.9\degr$ and $\approx 0.7\degr$ from the invariable plane, respectively. All this suggests that inner systems that are dynamically `cold' correlate with `cold' outer systems.

In order to test this correlation further, it is important to know the mutual inclinations between the inner planets and the outer planets. From our analysis, we find that $\pi$ Men and HAT-P-11 both have giant planets that are highly misaligned with their single transiting planets. The eccentricities of the giant planets in $\pi$ Men and HAT-P-11 are both high ($\gtrsim0.6$) as well. Although statistically insignificant as yet ($N=2$), these results do support the hypothesis that dynamically hot outer systems play a role in sculpting their inner systems.

Measuring the obliquity angle in systems with mutually misaligned planets would be another a strong test for correlations between inner and outer systems. As we have shown with $\pi$ Men and HAT-P-11, nodal precession alone can generate large obliquities \citep[see also][]{yee_hat-p-11_2018}. Indeed, our measurement of $54\degr < \Delta I < 126\degr$ ($1\sigma$) between HAT-P-11 b and c can explain the $\approx 100\degr$ obliquity of HAT-P-11. Therefore, we expect that inner and outer systems with large mutual inclinations between them will also tend to have large obliquities between their inner planets and their host stars. In fact, \citet{morton_obliquities_2014} found evidence that obliquities of stars with single transiting planets are higher than those with multiple transiting planets at the $\approx 2\sigma$ level, supporting a correlation between single transits and larger obliquities.

The eccentricities of the inner planets can be used as another proxy for the level of excitation in the inner system. HAT-P-11 b has a non-negligible eccentricity of $0.22\pm0.03$ \citep{yee_hat-p-11_2018}, and our simulations show that $\pi$ Men c could have a moderate eccentricity due to secular perturbations from $\pi$ Men b, although tidal dissipation may damp away the free eccentricity over Gyr time scales \citep[see \S\ref{sec:max_e}; a tentative $1\sigma$ upper bound of 0.3 was found in][]{huang_tess_2018}. \citet{van_eylen_orbital_2019} showed that the observed eccentricities of single-transiting systems are statistically higher than those of multiple-transiting systems: when modelled with a Rayleigh distribution, the scale parameter ($\sigma$) is $0.24\pm0.04$ for singles and $0.06\pm0.01$ for multis. This suggests that moderate inner eccentricities do go hand-in-hand with higher mutual inclinations or lower intrinsic multiplicities of the inner system, both of which reduce the transit multiplicity \citep[see also][]{huang_dynamically_2017}.

Yet another approach has been taken by \citet{zhu_super_2018}, who used the metallicity of the host star as a proxy for CJ occurrence, following their well-established relationship \citep{fischer_planet-metallicity_2005-1}. Using data from RV follow-up surveys of {\it Kepler} systems, they found that there is a deficit of high-multiplicity systems around high-metallicity host stars at the $93$ per cent confidence level. This serves as another line of evidence that systems with outer CJs might have disrupted their inner systems.

In summary, there is an emerging link between the following: hot outer systems (eccentric and inclined CJs) $\Rightarrow$ hot inner systems $\rightarrow$ low transit multiplicities $\&$ high obliquities $\&$ high inner eccentricities. Our measurements and analysis of $\Delta I$ in $\pi$ Men and HAT-P-11 not only support this correlation, but also demonstrate that nodal precession is a plausible explanation for part of the correlation, namely that between misaligned inner and outer systems and high obliquities between the star and the inner system. On the other hand, there is also tentative evidence for the opposite correlation: cold outer systems or absence of outer systems $\Rightarrow$ cold inner systems $\rightarrow$ high transit multiplicities $\&$ low obliquities $\&$ low inner eccentricities. More multi-planet systems with measurements of eccentricities, obliquities, and inclinations are needed to further prove or disprove these correlations. For example, dedicated RV monitoring of {\it Kepler} systems in search for outer CJs would help increase the sample size \citep[e.g.][]{mills_long-period_2019}.

\subsection{Planet formation and evolution in large $\Delta I$ systems}\label{sec:diss_formation}
What causes $\Delta I$ to be large in $\pi$ Men and HAT-P-11, and what does this mean for the initial stages of planetary formation and subsequent evolution in these systems? In \S\ref{sec:geom_overview}, we outlined three possible configurations of $\Lone$,  $\Ltwo$, and $\Lstar$. To discuss formation scenarios, we must also consider the orientations of the protoplanetary disc and any present-day debris disc, which we refer to as $\Lppd$ and $\Ldd$, respectively.

Both $\pi$ Men and HAT-P-11 are evolved systems with ages of several Gyr. There is no known debris disc in HAT-P-11, but $\pi$ Men hosts a faint debris disc (fractional luminosity of $\sim$ \SI{1.6e-6}, blackbody radius $\sim 30-170$ au) detected from far-infrared excess with {\it Herschel} \citep{sibthorpe_analysis_2018}. While over a hundred debris discs have been imaged in scattered light and thermal emission, the faintness of the $\pi$ Men disc makes it very challenging to image; a resolved image of this disc would require long integration times, for example with ALMA, or improved sensitivity of future instruments such as SPICA \citep{roelfsema_spica_2018}. If imaged, the orientation and structure of the disc would provide significant benefits in our understanding of the formation of $\pi$ Men, as we discuss below.

In the following, we shall refer to initial orientations with the subscript `i', and long-term orientations with the subscript `lt'. We use the three general cases outlined in \S\ref{sec:geom_overview} to describe long-term configurations. The first two subsections below (\S\ref{sec:giant_scat} and \S\ref{sec:self_excite}) focus on planet-planet interactions to drive the large observed $\Delta I$ between the inner and outer planets of $\pi$ Men and HAT-P-11. In these subsections, we assume that at the initial stage of planet formation, $\Lppdi \approx \Lstari \approx \Lonei \approx \Ltwoi$, i.e. the system was initially flat before the onset of later perturbations. When we consider misaligned protoplanetary discs in the third subsection (\S\ref{sec:misaligned_ppd}), this assumption is dropped.

\subsubsection{Planet-planet scattering in the outer system}\label{sec:giant_scat}
Given the high eccentricity of the outer planets in $\pi$ Men and HAT-P-11 ($e\gtrsim0.6$), it is plausible that they originated in groups of closely packed giant planets that underwent unstable dynamical interactions. Previous studies show that systems with three or more giant planets that undergo dynamical encounters after disc dissipation can lose all but one remaining planet, which ends up with a high eccentricity and high inclination relative to the initial plane of the disc \citep[e.g.][]{chatterjee_dynamical_2008, juric_dynamical_2008}. As noted before, we assume the system was entirely flat before the outer planet was affected by dynamical scattering. After scattering, we get $\Ltwolt \ne \Lstarlt$.

After the outer planet becomes misaligned with the stellar spin axis, it will inevitably influence the inner orbit, which is initially aligned with the star. If the inner planet formed in situ, where large amplitude oscillations in $e_1$ and $I_1$ are suppressed, nodal precession will cause $\boldsymbol{\hat{L}}_{1}$ to evolve with time and therefore rapidly lose memory of any initial alignment with $\Lstar$ (see \S\ref{sec:geom_overview}). The stellar spin axis also precesses, but for both $\pi$ Men and HAT-P-11, it does so in ways that maintain the misalignment between $\boldsymbol{\hat{L}}_{1}$ and $\Lstar$. (see \S\ref{sec:weakcoup_pimen} for details). The system would therefore be observed as case III most of the time ($\Lonelt \ne \Lstarlt \ne \Ltwolt$). 

On the other hand, if the inner planet formed at a larger semi-major axis, where the GR precession rate is slow, secular perturbations from the outer planet might have brought the inner eccentricity to extreme values ($\gtrsim 0.9$), causing the planet to tidally migrate until GR starts to suppress the large oscillations in $e_1$ and $I_1$. During these oscillations, the inclination of the inner orbit changes significantly, which again causes $\Lonelt \ne \Lstarlt$ and gives case III. In this scenario, however, it is unclear why some planets end up in sub-day orbits \citep[i.e. forming ultra-short period planets as in][]{petrovich_ultra-short-period_2019-1}, while others planets like $\pi$ Men c and HAT-P-11 b end up in orbits that are relatively stable against further orbital decay.

After scattering occurs between the giant planets, we expect the debris disc to trace the orientation of the protoplanetary disc and thus be initially misaligned with the orbit of the outer planet (i.e. $\Lppdi \approx \Lddi \ne \Ltwoi$). However, the eccentric giant planet would then act on the less massive exterior debris disc and alter its shape and orientation via secular interactions. \citet{pearce_dynamical_2014} modelled these interactions and found a few different stable outcomes depending on the planet eccentricity and initial planet-disc mutual inclination, $\Delta I_{\rm{pd, i}}$. The time scale for the debris disc in $\pi$ Men to evolve into these outcomes is on the order of 10 Myr assuming a disc radius of $\sim100$ au (using eq. 17 of \citealt{pearce_dynamical_2014}), well below the system age. There is a critical $\Delta I_{\rm{pd, i}}$ that separates different outcomes, which we estimate to be $\sim 30\degr$ for $\pi$ Men b (using eq.13, originally from \citealt{farago_high_2010}). Given $I_{\rm{2,sky}} = 41-65\degr$, and $I_\star = 63-85\degr$ for $\pi$ Men, the critical $\Delta I_{\rm{pd, i}}$ might or might not be attained in this system.

We summarize the different cases found in \citet{pearce_dynamical_2014} (see their fig.3 for visual illustrations). Below the critical $\Delta I_{\rm{pd, i}}$, the disc would become aligned with the planet's orbit but puffed up vertically, i.e. $\Lddlt \approx \Ltwolt$. In addition, the disc would become eccentric and apsidally aligned with the planetary orbit, so it would appear elongated along the line of apsides. Above the critical angle, there are two different possibilities. If $\Delta I_{\rm{pd, i}}$ is not too high, the disc would form a hollow bell-shaped structure that encapsulates the planet's orbit, becoming more like a cloud than a disc. In this case, we still have $\Lddlt \approx \Ltwolt$, but the disc would have distinct features such as an overdensity at the apocentre. In a more extreme case, if $\Delta I_{\rm{pd, i}} \gtrsim 60\degr$, the disc would become orthogonal to the planetary orbit (i.e. $\Lddlt$ and $\Ltwolt$ are maximally misaligned). Such polar discs have been observed around binary stars \citep{kennedy_99_2012, kennedy_circumbinary_2019}. We might expect that it would be difficult for planet-planet dynamical scattering to excite such extreme inclinations, however, so the last possibility may be unlikely for $\pi$ Men. Overall, we would expect $\Lddlt \approx \Ltwolt$, unless $\Delta I_{\rm{pd, i}}$ is very large in which case a polar debris disc forms. The exact structure of the disc would provide constrains on $\Delta I_{\rm{pd,i}}$, which sheds light on the early conditions of the system.

\subsubsection{Self-excitation of the inner system}\label{sec:self_excite}
It is also possible that the giant planet and the stellar spin axis are currently aligned ($\Lstarlt \approx \Ltwolt$), but the inner planet is misaligned with both (case II). Like in the previous case, assume that both planets formed in a flat protoplanetary disc and were originally co-planar. If the inner planet had multiple close neighbours, it might have experienced a period of dynamical instability that led to the ejection or collision of those neighbouring planets. In this process, the inner planet acquires inclination, causing $\Lonelt \ne \Lstarlt \approx \Ltwolt$. This is similar to the process in the previous section, but between lower-mass inner planets. These excitations have been modelled during the period of in situ assembly, and are able to produce diverse systems from different initial conditions of the protoplanetary disc \citep{dawson_correlations_2016, moriarty_kepler_2016}. As noted in \S\ref{sec:geom_overview}, an important feature of case II is the we expect $\psi \approx \Delta I$.

After self-excitation of the inner system, we expect the outer debris disc to remain aligned with the outer planet, as it traces the orientation of the protoplanetary disc (i.e. $\Lppdi \approx \Lddlt \approx \Ltwolt$). If the system evolved in this way to give case II, we would expect the debris disc to be much thinner than the ``re-aligned'' discs described in \S\ref{sec:giant_scat}, with a scale height of $\approx$ 0.03-0.1 similar to other imaged discs \citep{matra_kuiper_2019, daley_mass_2019}. Therefore, measuring the disc scale height would help determine whether the debris disc was initially misaligned to the giant planet and later re-aligned, or always aligned to the giant planet.

However, one caveat to this scenario is that gravitational interactions between these less massive planets have difficulties exciting the inclinations by more than $\sim 10\degr$, and are perhaps insufficient to explain the mutual inclinations that we observe in $\pi$ Men and HAT-P-11. In addition, this scenario does not explain the observed large eccentricities of the outer planets. It might then be more likely that planet-planet scattering took place in both the inner system and the outer system, with the outer system simultaneously influencing the evolution of the inner system. These interactions can easily misalign the planets with respect to each other and their star, again reducing the configuration to case III ($\Lonelt \ne \Lstarlt \ne \Ltwolt$). The outer planet will tend to re-align the debris disc to its orbit, so we expect the same disc outcomes as described in \S\ref{sec:giant_scat}.

\subsubsection{Broken and misaligned protoplanetary discs}\label{sec:misaligned_ppd}
In the two previous cases, the planets form in a flat protoplanetary disc aligned with the star, but subsequently evolve to obtain large mutual inclinations. Here, we consider the alternative possibility that the planets formed in a broken protoplanetary disc with misaligned inner and outer components.

Simulations show that a misalignment between the inner and outer discs could be caused by a misaligned gas giant companion \citep{zhu_inclined_2019, nealon_scattered_2019} or a misaligned binary star system \citep[e.g.][]{facchini_signatures_2018}. Alternatively, it is possible that there are additional physical effects that break and misalign the discs. Observationally, the mutual inclinations between inner and outer disc components have been measured in some systems and range between $30\degr$ to $80\degr$ \citep[e.g.][]{marino_shadows_2015, loomis_multi-ringed_2017, min_connecting_2017, walsh_co_2017, casassus_inner_2018}. If the discs are misaligned by giant companions, we expect the companion's orbit to be inclined relative to either disc by about half or more their mutual inclination due to nodal precession \citep{zhu_inclined_2019}. This implies companions inclined to their discs by $>15-40\degr$ in the currently observed sample, which is comparable in size to the misalignment between $\pi$ Men b and $\pi$ Men c. 

In the following, we assume that the giant planets caused primordial misalignments in the protoplanetary discs around $\pi$ Men and HAT-P-11. Therefore, we have $\boldsymbol{\hat{L}}_{\rm{in,ppd,i}} \ne \boldsymbol{\hat{L}}_{\rm{out,ppd, i}} \ne \Ltwoi$, where `in' and `out' refer to the inner and outer disc components. As the giant planet induces precession of the inner and outer discs, both discs become misaligned with the stellar spin axis, which we assume is also misaligned with the giant planet. If the inner planet formed in the inner disc, then $\boldsymbol{\hat{L}}_{\rm{in,ppd,i}} \approx \Lonei$, the only initial alignment in the system. The system therefore starts off with $\Lonei \ne \Lstari \ne \Ltwoi$. Further secular interactions between the inner and outer planet only tend to maintain these misalignments, so we get case III, $\Lonelt \ne \Lstarlt \ne \Ltwolt$. Under this picture, the giant planets in $\pi$ Men and HAT-P-11 would serve as fossil records of misaligned protoplanetary discs. It is unclear, however, how the giant planets were misaligned with the protoplanetary discs in the first place, although formation in a dense stellar cluster \citep{bate_diversity_2018} or early phase planet-planet scattering in the outer system may offer possible avenues.

Assuming the debris disc traces the orientation of $\boldsymbol{\hat{L}}_{\rm{out,ppd,i}}$, it would be initially misaligned with the giant planet, but later reshaped by it. We would expect to observe the same disc structure as described in the end of \S\ref{sec:giant_scat}. However, if the disc misalignment was caused by another physical process unrelated to the giant planet, the giant planet may have formed coplanar to the outer disc. In this case, the debris disc would be vertically thin and aligned with the $\Ltwo$ from the beginning. As in the inner system self-excitation scenario (\S\ref{sec:self_excite}), this would produce case II ($\Lonelt \ne \Lstarlt \approx \Ltwolt$).

\subsubsection{Summary}
After considering the dynamics and possible formation scenarios, it appears that case III ($\Lonelt \ne \Lstarlt \ne \Ltwolt$) is the most likely long-term configuration for the planets and host star in $\pi$ Men and HAT-P-11. Case I ($\Lonelt \approx \Lstarlt \ne \Ltwolt$) is unlikely because nodal precession renders alignments between $\Lone$ and $\Lstar$ rare (see \S\ref{sec:nod_prec_effects}), while case II ($\Lonelt \ne \Lstarlt \approx \Ltwolt$) requires dynamical interactions between the inner planets alone to generate the large measured $\Delta I$ values, or additional physics to misalign protoplanetary discs.

Indeed, observations of the stellar obliquity and stellar spin orientation in HAT-P-11 point to case III for this system (see \S\ref{sec:stellar_spin}). For $\pi$ Men, the stellar spin axis is consistent with being aligned with the inner planet, the outer planet, or neither. If a large obliquity is measured for $\pi$ Men, this would definitively rule out case I. However, as case I is fortuitous, the long-term configuration should be either case II or case III. In terms of formation, case III would be the result of either violent scattering between giant planets (\S\ref{sec:giant_scat}), or formation in a misaligned protoplanetary disc torqued by an initially misaligned giant planet (\S\ref{sec:misaligned_ppd}). On the other hand, case II would imply either self-excitation of the inner system (\S\ref{sec:self_excite}) or a misaligned protoplanetary disc that was not produced by the giant planet. Distinguishing between these two cases for $\pi$ Men would require a better measurement of $I_\star$ through asteroseismology. In addition, a telltale sign of case II would provided be if $\psi \approx \Delta I$, which could be tested with measurements of the Rossiter-McLaughlin effect.

We can also differentiate between different formation scenarios based on the structure and orientation of the $\pi$ Men debris disc. As noted in \S\ref{sec:self_excite}, if $\Lddlt \approx \Ltwolt$ and the disc has a small scale height of $\lesssim 0.1$, then the formation scenarios that produce case II would be supported. Alternatively, if we find that $\Lddlt \approx \Ltwolt$ and the disc is vertically extended with a large scale height, the formation scenarios that give case III are more likely. By further characterising the disc structure, we can even determine the initial mutual inclination between the planet and the debris disc ($\Delta I_{\rm{pd,i}}$) using results from \citet{pearce_dynamical_2014}. We expect the debris disc to be vertically extended and elongated along the line of apsides for $\Delta I_{\rm{pd,i}} \lesssim 30\degr$, and hollow and bell-shaped with an overdensity at the apocentre for $\Delta I_{\rm{pd,i}} \gtrsim 30\degr$. Lastly, observing a debris disc orthogonal to $\Ltwolt$ would be evidence that $\Delta I_{\rm{pd,i}} \gtrsim 60\degr$. Imaging the debris disc in $\pi$ Men would therefore provide a significant advantage in our understanding of the initial conditions of planet formation in this system.

\section{Conclusions}\label{sec:conclusion}
In conclusion, we have measured the mutual inclination between the cold Jupiter $\pi$ Men b and the super Earth $\pi$ Men c using a combination of transit, RV, and astrometric data. Our astrometric data comes from proper motion anomalies between {\it Gaia} and {\it Hipparcos} measurements, and we validate our method by applying it to the brown dwarf companion HD 4747. We find that $49\degr < \Delta I < 131\degr$ (1$\sigma$), $28\degr < \Delta I < 152\degr$ (2$\sigma$), and $9\degr < \Delta I < 171\degr$ (3$\sigma$) between the $\pi$ Men planets. We study the dynamics in $\pi$ Men, and find that despite the high mutual inclinations, potentially large oscillations in $e_1$ and $I_1$ are suppressed by GR precession. On the other hand, nodal precession becomes an important secular phenomenon in the system, and can generate large stellar obliquities and shift the inner planet out of a transiting configuration. We compare the dynamics in $\pi$ Men to that in HAT-P-11, where we find that $54\degr < \Delta I < 126\degr$ at the $1\sigma$ level, although the astrometric data is less significant. The large $\Delta I$ we measure in HAT-P-11 supports nodal precession as the explanation of its $\approx 100\degr$ sky-projected obliquity, first proposed by \citet{yee_hat-p-11_2018}.

The large $\Delta I$ in $\pi$ Men and HAT-P-11 support the idea that dynamically hot outer systems have shaped their inner systems, with a range of possible effects including (i) reducing the transit multiplicity, (ii) increasing the eccentricity of inner planets, (iii) increasing the stellar obliquity. In contrast to the population of flat {\it Kepler} systems with multiple transiting planets, where the mutual inclinations are on the order of a few degrees, $\pi$ Men and HAT-P-11 seem to have experienced more violent dynamical histories. With more TESS observations, long-term RV monitoring, and future {\it Gaia} epoch astrometry, the links between inner and outer systems can be tested with larger sample sizes. Our measurements also shed light on processes that govern planet formation and evolution. For example, the large mutual inclinations in $\pi$ Men and HAT-P-11 suggest a history of chaotic planet-planet scattering in the outer system, or formation in misaligned protoplanetary discs. Imaging of the debris disc in $\pi$ Men would help place further constraints on which formation scenario is more likely. These formation processes, as well as latter day secular effects such as nodal precession, tend to shift the orbits out of alignment with each other and the stellar spin axis.

\section*{Acknowledgements}
We thank the anonymous referee and the second referee Billy Quarles for helpful comments on the manuscript. We are grateful to Grant Kennedy for pointing our attention to the proper motion anomaly data. JX acknowledges funding from the Downing Scholarship of Pomona College and Downing College, Cambridge. JX also thanks Pierre Kervella, Timothy Brandt, Chelsea Huang, Kento Masuda, Bonan Pu, Steven Young, Joshua Winn, Samuel Yee, and Jingwen Zhang for helpful discussions. Simulations in this paper made use of the REBOUND code which is freely available at http://github.com/hannorein/rebound.

This work has made use of data from the European Space Agency (ESA) mission {\it Gaia} (\url{https://www.cosmos.esa.int/gaia}), processed by the {\it Gaia} Data Processing and Analysis Consortium (DPAC, \url{https://www.cosmos.esa.int/web/gaia/dpac/consortium}). Funding for the DPAC has been provided by national institutions, in particular the institutions participating in the {\it Gaia} Multilateral Agreement.

\section*{Data Availability}
The data for proper motion anomalies used this article are available in \citet{brandt_erratum_2019}, at \url{https://doi.org/10.3847/1538-4365/ab13b2}. The radial velocity data used are available in \citet{tal-or_correcting_2019} (\url{http://vizier.u-strasbg.fr/viz-bin/VizieR?-source=J/MNRAS/484/L8}), \citet{gandolfi_tesss_2018} (\url{http://cdsarc.u-strasbg.fr/viz-bin/qcat?J/A+A/619/L10}), \citet{trifonov_public_2020} (\url{https://cdsarc.unistra.fr/viz-bin/cat/J/A\%2bA/636/A74}), and \citet{yee_hat-p-11_2018} (\url{https://cdsarc.unistra.fr/viz-bin/cat/J/AJ/155/255}). 

\bibliographystyle{mnras}
\bibliography{pimen}

\appendix

\section{Transforming from sky plane to invariable plane} \label{appendixA}
To transform between the two reference planes, we first specify the location of the invariable plane with respect to the sky plane, which is set by two angles $I'$ and $\Omega'$. $I'$ and $\Omega'$ can be computed by considering the conservation of angular momentum. Specifically,
\begin{equation}
    \begin{aligned}
    L_x &= L_{\rm{tot}} \sin{I'} \sin{\Omega'} = \sum_{j}^{1,2}{L_j \sin{I_{j,sky}} \sin{\Omega_{j,sky}}}, \\
    L_y &= -L_{\rm{tot}} \sin{I'} \cos{\Omega'} = -\sum_{j}^{1,2}{L_j \sin{I_{j,sky}} \cos{\Omega_{j,sky}}}, \\
    L_z &= L_{\rm{tot}} \cos{I'} = \sum_{j}^{1,2}{L_j \cos{I_{j,sky}}},
    \label{eq:Lxyz}
    \end{aligned}
\end{equation}
where $L_j$ is the angular momentum of the $j$th planet, which is proportional to $m_j \sqrt{a_j (1-e_j^2)}$ if we ignore the spin angular momentum. $m_j$, $a_j$, and $e_j$ are the mass, semi-major axis, and eccentricity of the $j$th planet, respectively. $L_{\rm{tot}}$ is the magnitude of the total angular momentum. Rearranging, we get
\begin{equation}
    \begin{aligned}
        I' &= \arccos\left(\frac{L_z}{L_{\rm{tot}}}\right), \\
        \Omega' &= \arctan\left(\frac{-L_x}{L_y}\right).
    \label{eq:Ip_Omega_p}
    \end{aligned}
\end{equation}
We note that for the $\pi$ Men and HAT-P-11 systems, the invariable plane is very close to the orbital plane of the outer planet, since it possesses most of the angular momentum in the system.

With the two angles $I'$ and $\Omega'$, and the three sky-projected orbital angles $\Omega_{\rm{sky}}$, $\omega_{\rm{sky}}$, and $I_{\rm{sky}}$, we can derive the respective $\Omega_{\rm{inv}}$, $\omega_{\rm{inv}}$, and $I_{\rm{inv}}$ using spherical trigonometry, which allows us to transform between the sky plane and the invariable plane. In terms of sky-projected angles, as well as $I'$ and $\Omega'$, we have
\begin{equation}
    \begin{aligned}
    I_{\rm{inv}} &= \arccos\left[\cos{I'} \cos{I_{\rm{sky}}} + \sin{I'} \sin{I_{\rm{sky}}} \cos{(\Omega_{\rm{sky}}-\Omega')}\right], \\
    \Omega_{\rm{inv}} &= \arctan\left[\frac{\sin{(\Omega_{\rm{sky}}-\Omega')}}{\cos{(\Omega_{\rm{sky}}-\Omega')}\cos{I'}-\sin{I'}/\tan{I_{\rm{sky}}}}\right], \\
    \omega_{\rm{inv}} &= \omega_{\rm{sky}} - \arcsin\left[\frac{\sin{I'}\sin{(\Omega_{\rm{sky}}-\Omega')}}{\sin{I_{\rm{inv}}}}\right].
    \label{eq:sky_inv}
    \end{aligned}
\end{equation}
These transformation equations satisfy the relations $I_{\rm{1,inv}}+I_{\rm{2,inv}} = \Delta I$ and $\lvert{\Omega_{\rm{1,inv}}-\Omega_{\rm{2,inv}}\rvert} = \pi$, which are always true in the invariable plane because the total angular momentum always points in the $z$ direction.

\section{Posterior distributions for $\pi$ Men} \label{appendixB}

\begin{figure*}
    \centering
    \includegraphics[width=1.0\linewidth]{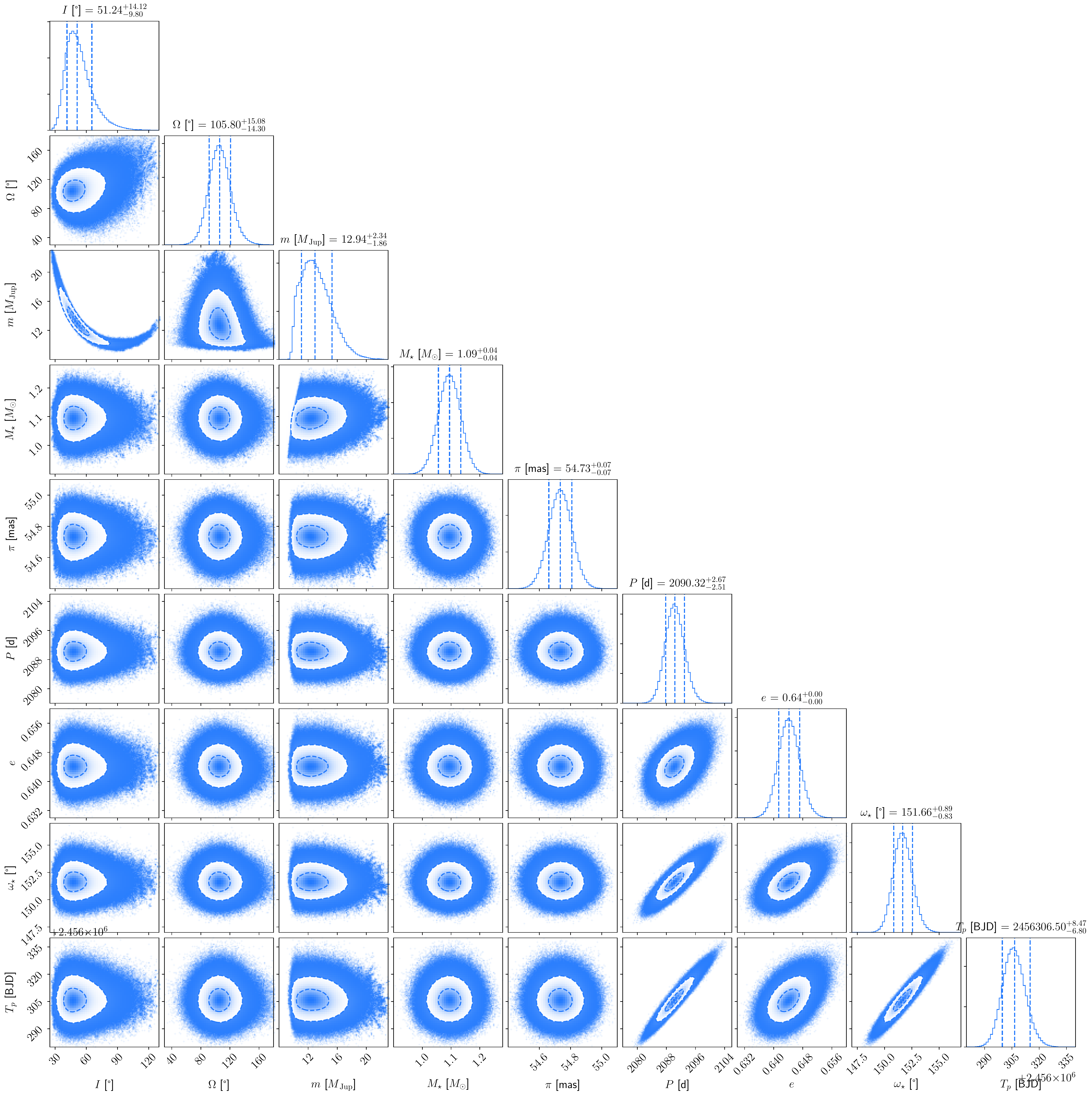}
    \caption{Target: $\pi$ Men b. Joint posterior distributions for the nine orbital parameters from our joint RV and PMa fits. Moving outward, the dashed lines on the 2D histograms correspond to $1\sigma$ and $2\sigma$ contours.}
    \label{fig:hd_corner}
\end{figure*}

\bsp	
\label{lastpage}
\end{document}